\documentclass[11pt]{scrartcl}

\usepackage[utf8]{inputenc}
\usepackage[T1]{fontenc}
\usepackage[english]{babel}

\usepackage{lmodern}
\usepackage{microtype}

\usepackage{amsmath,amssymb,amsfonts}

\usepackage{graphicx}
\usepackage{hyperref}

\usepackage{geometry}
\usepackage{float}
\usepackage{xcolor}

\usepackage[format=plain,
            singlelinecheck=false,
            indention=0em,
            labelfont=bf]{caption}

\usepackage[symbol]{footmisc}

\newcommand{\email}[1]{\footnote{\texttt{#1}}}

\begin{document}

\begin{center}
    {\Large \textbf{Nonlinear dynamics in Horndeski gravity: a renormalized approach to effective gravitational coupling} \\[1.2em]}
    {\large Luca Amendola$^{1, }$\email{l.amendola@thphys.uni-heidelberg.de}, Carla Bernal$^{2, }$\email{carlabernalb@gmail.com (corresponding author)}, Radouane Gannouji$^{2, }$\email{radouane.gannouji@pucv.cl} \\[0.5em]}
    {\small $^1$Institut für Theoretische Physik, Universität Heidelberg, Philosophenweg 16, 69120 Heidelberg, Germany \\ 
    $^{2}$Instituto de Física, Pontificia Universidad Católica de Valparaíso Av. Brasil 2950, Valparaíso, Chile}
\end{center}

\vspace{1em}

\begin{abstract}
This paper develops a renormalized perturbation theory framework for nonlinear structure formation in a broad class of modified gravity models that exhibit Vainshtein screening, with a focus on a viable subclass of Horndeski theories. We extend earlier perturbative methods, originally applied to DGP model, to construct a self‐consistent treatment that captures both the linear modifications to gravity at large scales and the nonlinear screening effects at small scales. In the framework, the response of the gravitational potential to matter density fluctuations is characterized by renormalized propagators, leading to the definition of a nonlinear (or renormalized) effective gravitational constant. The paper details several numerical strategies to compute this renormalized gravitational constant. Numerical examples illustrate how the effective gravitational constant evolves with scale and redshift. These results are key to accurately predicting cosmological observables such as the matter power spectrum and bispectrum in modified gravity scenarios.
\end{abstract}

\section{Introduction}

The late-time accelerated expansion of the Universe remains a central puzzle in modern cosmology. While the standard $\Lambda$CDM model associates this phenomenon to a cosmological constant, many cosmologists have focused on the possibility that gravity itself could be modified on large scales. Among the broad classes of modified theories \cite{Clifton:2011jh,Joyce:2014kja}, Horndeski's theory \cite{Horndeski:1974wa,Deffayet:2009wt,Kobayashi:2011nu} and its generalizations stand out as the most general scalar-tensor theory yielding second-order equations of motion and thus free of Ostrogradski-type instabilities, which arises when higher-order time derivatives introduce extra, potentially ghost-like degrees of freedom. Therefore these theories bring higher order derivatives in a consistent way. Indeed, the Galileon theory has the remarkable property that certain higher-derivative operators remain exactly non-renormalized by quantum corrections \cite{Luty:2003vm,Goon:2016ihr}. Essentially, the shift symmetry forces loop corrections to vanish or be re-expressible in terms of lower-derivative terms. Unfortunately, once the Galileon field is coupled to gravity, this exact symmetry is broken and so one loses the strict non-renormalization. However, there are special subsets of these theories that preserve a remnant of the Galileon symmetry and the quantum corrections to the higher-derivative terms may remain suppressed \cite{Pirtskhalava:2015nla,Santoni:2018rrx}, preserving some of the non-renormalization arguments at least to some order or around special backgrounds.

These theories exhibit a rich phenomenology, particularly in their ability to evade local gravitational constraints through screening mechanisms. A notable example is the Vainshtein mechanism, initially developed in the context of massive gravity \cite{Vainshtein:1972sx,Boulware:1972yco} but also relevant to braneworld models such as the DGP model. In these scenarios, modifications to the gravitational force are suppressed in regions of high density or strong curvature, ensuring that General Relativity is effectively restored at small scales. However, the mechanism is often studied in simplified settings, such as spherical symmetry \cite{Babichev:2013usa}. A realistic macroscopic object, however, consists of many point particles, each with its own Vainshtein radius, making the determination of the effective Vainshtein radius of the entire object nontrivial. This challenge, known as the "elephant problem" \cite{Clifton:2011jh}, is crucial for understanding structure formation.

Early investigations of how Vainshtein screening impacts structure formation were carried out, for instance, in the DGP scenario \cite{Scoccimarro:2009eu}, where a non-linear perturbative approach was developed suited to scales where screening becomes important. In \cite{Brando:2023fzu}, the authors improved their N-body code using the non-linear effective gravitational constant of \cite{Scoccimarro:2009eu}. Our paper generalizes the methods of \cite{Scoccimarro:2009eu} to a class of Horndeski-like models endowed with Vainshtein screening. Specifically, we formulate a renormalized perturbation theory framework designed to capture both the large-scale, linearized modifications to gravity and the small-scale screening behavior in a self-consistent way. 

The remainder of this paper is organized as follows. In Sec. \textbf{2}, we briefly review the viable subclass of Horndeski theories arising after constraints from GW170817 and introduce the background equations. In Sec. \textbf{3}, we present the quasi-static limit for scalar and metric perturbations in the non-linear regime, emphasizing the role of Galileon-like interactions for Vainshtein screening. Sec. \textbf{4}, \textbf{5}, \textbf{6} and \textbf{7} describe the renormalized perturbation theory approach and applies it to compute non-linear responses of the gravitational potential. In Sec. \textbf{9} and \textbf{10} we show numerical results for a representative model and discuss the impact on cosmological observables. Finally, we summarize our conclusions and outline future research directions in Sec. \textbf{11}.

\section{Viable Horndeski's theories}

Among scalar field-modified gravity theories, Horndeski models have earned a prominent position. Various papers have studied these models and their applications in cosmology or black hole physics. One particularly interesting result stems from the observation of the gravitational-wave event GW170817 and the gamma-ray burst 170817A \cite{LIGOScientific:2017zic}, which constrained the full Horndeski theories to a viable subclass \cite{Creminelli:2017sry,Sakstein:2017xjx,Ezquiaga:2017ekz}:
\begin{align} 
\label{action}
\mathcal{S} = \int {\textnormal d}^4 x \sqrt{-g} \Bigl[\frac{F(\phi)}{16\pi G} R + K(\phi, X) - G(\phi, X) \square \phi\Bigr], 
\end{align}
where $\phi$ is the scalar field and $X = -(\partial_\mu \phi)^2/2$ is the kinetic term. The functions $K$ and $G$ are arbitrary. This model corresponds to the generalized Kinetic Gravity Braiding introduced in \cite{Deffayet:2010qz}.

This constraint has, of course, been challenged in the literature \cite{deRham:2018red,Amendola:2018ltt}, but we will adopt a conservative approach and consider these models as the viable subclass. However, this constraint does not guarantee that the model is entirely safe. For instance, generic Horndeski theories often lack a well-posed Cauchy problem, meaning that solutions may cease to exist after some time for generic initial conditions. The equations lose hyperbolicity \cite{Papallo:2017qvl,Ripley:2019hxt,Ripley:2019irj,Ripley:2019aqj} in the strong regime, where higher-derivative effects dominate. In the weak regime, however, this subclass of theories remains viable \cite{Kovacs:2020ywu}.

From an effective field theory perspective, the weak regime is the only domain where these theories can be trusted, as additional corrections are expected to become significant in the strong regime. Therefore, we will study the class of theories (\ref{action})

The Einstein equations are
\begin{align}
& F \Bigl(R_{\mu\nu}-\frac{1}{2}R g_{\mu\nu}\Bigr)+g_{\mu\nu} \Box F-\nabla_{\mu\nu} F =8\pi G \Bigl(T_{\mu\nu}+T_{\mu\nu}^{(K)}+T_{\mu\nu}^{(G)}\Bigr)\\
& \nabla_\mu J^\mu = K_{,\phi}+\phi^{,\mu}\nabla_\mu G_{,\phi}+\frac{F'(\phi)}{16\pi G}R
\end{align}
with
\begin{align}
T_{\mu\nu}^{(K)} & = K g_{\mu\nu}+K_{,X} \phi_{,\mu}\phi_{,\nu} \\
T_{\mu\nu}^{(G)} & = - 2 G_{,\phi} \Bigl(g_{\mu\nu} X+\phi_{,\mu}\phi_{,\nu}\Bigr)-G_{,X}\Bigl(g_{\mu\nu}\phi^{,\alpha}\phi_{,\alpha\beta}\phi^{,\beta}-2\phi_{,\alpha(\mu}\phi^{,\alpha}\phi_{,\nu)}+\phi_{,\mu}\phi_{,\nu}\Box\phi\Bigr)\\
J_\mu &= G_{,\phi}\phi_{,\mu}+\nabla_\mu G+G_{,X}\phi_{,\mu}\Box\phi-K_{,X}\phi_{,\mu}
\end{align}
Assuming that the Universe is described by a flat FLRW metric ${\textnormal d}s^2=-{\textnormal d}t^2+a^2(t) {\textnormal d}\mathbf{x}^2$ and that only pressure-less matter is present, the Einstein equations reduce to
\begin{align}
& 3 H^2 F+3 H \dot{\phi} F'
 = 8\pi G \Bigl(\rho_m- K+2 X K_{,X}
-2 X G_{,\phi}+6 X \dot{\phi} H G_{,X}\Bigr)\\
& \left(3 H^2+2 \dot{H}\right) F 
+(\ddot{\phi}+2 H \dot{\phi}) F' 
+2 X F''
= 8\pi G \Bigl(2 X\left(G_{,\phi}+\ddot{\phi} G_{,X}\right)-K\Bigr)
\end{align}
while the generalized Klein-Gordon equation is
\begin{align}
\dot J+3 H J= K_{,\phi}-2 X\left(G_{,\phi \phi}+\ddot{\phi} G_{, \phi X}\right) +3\left(2 H^2+\dot{H}\right) \frac{F'}{8\pi G}
\end{align}
where
\begin{align}
    J\equiv J^0=\dot{\phi} K_{,X}+6 H X G_{,X}-2 \dot{\phi} G_{,\phi}
\end{align}

\section{Perturbations with sufficient nonlinearities}

In this section, we perturb the spacetime while retaining relevant non-linearities. We work in the Newtonian gauge and adopt the quasi-static approximation, even though this may be a somewhat restrictive assumption \cite{Sawicki:2015zya}:
\begin{align} 
\mathrm{d} s^2 = -(1 + 2\Phi) \mathrm{d} t^2 + a^2(1 - 2\Psi) \mathrm{d} \mathbf{x}^2, 
\end{align}
with the scalar field perturbation given by $\phi(t,\mathbf{x}) = \phi(t) + \delta \phi(t,\mathbf{x})$.

Although the perturbations are assumed to be small, their spatial derivatives can become significant on sufficiently small scales. For this reason, we retain all spatial derivatives while neglecting time derivatives under the quasi-static approximation. Non-linear relativistic corrections are not considered, so we limit the metric perturbations to first order in $(\Phi, \Psi)$. However, scalar field perturbations $\delta\phi$ could, in principle, include higher-order contributions.

In summary, we assume $\delta\phi$ is small, but its spatial derivatives can be large on small scales. Therefore, terms such as $(\delta\phi)^n \delta\phi_{,ij} \ll \delta\phi_{,ij}$ and $\delta \phi_{,i} \delta \phi_{,j} \ll \delta \phi_{,ij}$ hold. Retaining only the relevant terms, we derive the following equation from the $(i,j)$ component with $i \neq j$:
\begin{align}
\partial_i\partial^j\Bigl(F(\phi)\Psi-F(\phi)\Phi-F'(\phi)\delta\phi\Bigr)=0
\end{align}
which implies 
\begin{align}
\Psi-\Phi=\frac{F'(\phi)}{F(\phi)}\delta\phi
\end{align}
Considering the $(0,0)$ equation, we find
\begin{align}
& F(\phi)\Delta\Psi +C_1\Delta \delta\phi = 4\pi G a^2\rho_m\delta_m
\end{align}
where
\begin{align}
C_1 &= 8\pi G X G_{,X}-\frac{F'}{2} = \frac{H F}{2\dot\phi} \alpha_B =\frac{F'}{2}\frac{\alpha_B}{\alpha_M}
\end{align}
We have introduced the parameters \cite{Bellini:2014fua}
\begin{align}
M_*^2 \equiv & \frac{F(\phi)}{8\pi G} \\ 
\alpha_{\mathrm{M}} \equiv & H^{-1}\frac{\mathrm{d} \ln M_*^2}{\mathrm{d} t}  = \frac{F'(\phi)\dot\phi}{H F(\phi)}\\ 
H^2 M_*^2 \alpha_K\equiv & 2 X\left(K_{,X}+2 X K_{,X X}-2 G_{,\phi}-2 X G_{,\phi X}\right)+ 12 \dot{\phi} X H\left(G_{,X}+X G_{,X X}\right) \\
H M_*^2 \alpha_{\mathrm{B}} \equiv & 2 \dot{\phi}\Bigl(X G_{,X}-\frac{F'(\phi)}{16\pi G}\Bigr)
\end{align}
Finally, the scalar field equation gives
\begin{align}
\label{eq:KG}
& F'(\phi)\Delta\Psi+C_1\Delta \Phi = \frac{4\pi G}{a^2}G_{,X}\Bigl((\Delta\delta\phi)^2-(\partial_{ij}\delta\phi)(\partial_{ij}\delta\phi)\Bigr)+C_2 F(\phi)\frac{H^2}{\dot\phi^2}\Delta \delta\phi
\end{align}
where
\begin{align}
C_2 &= -\frac{1}{2}K_{,X}+G_{,\phi}-2H\dot\phi G_{,X}-\ddot \phi G_{,X}-XG_{,\phi X}-X\ddot \phi G_{,XX}\\
& = \frac{3}{2}\Omega_m-\alpha_M-\frac{\alpha_B}{2}(1+\alpha_M)-\frac{\dot\alpha_B}{2 H}-\frac{\dot H}{H^2}\Bigl(\frac{\alpha_B}{2}-1\Bigr)\\
& \equiv \alpha_M^2-\Bigl(\alpha_M+\frac{\alpha_B}{2}\Bigr)^2-\frac{c_s^2}{2}\Bigl(\alpha_K+\frac{3}{2}\alpha_B^2\Bigr)
\end{align}
The density contrast is defined as $\delta_m = \delta \rho_m / \rho_m$, with $\Omega_m = 8\pi G \rho_m / (3F H^2)$, and $c_s$ representing the sound speed of the scalar mode \cite{Bellini:2014fua}, which depends only on the $\alpha$ parameters and the Hubble function. Our results are consistent with those reported in \cite{Kimura:2011dc}.

It is evident from eq.(\ref{eq:KG}) that the Galileon-type terms, specifically $G_{,X} \neq 0$, are crucial for introducing non-linearities and, consequently, enabling the Vainshtein mechanism. These equations can be expressed as
\begin{align}
& \Psi-\Phi=\alpha_M Q \\
& \Delta\Psi +\frac{\alpha_B}{2}\Delta Q = \frac{3}{2}a^2H^2\Omega_m\delta_m\\
& \alpha_M \Delta\Psi+\frac{\alpha_B}{2}\Delta \Phi = \frac{\alpha_B+\alpha_M}{2a^2H^2}\Bigl((\Delta Q)^2-(\partial_{ij}Q)(\partial_{ij}Q)\Bigr)+C_2 \Delta Q
\end{align}
where we introduced the variable \cite{Kimura:2011dc} $Q=H\delta\phi/\dot\phi$. These equations reduce to
\begin{align}
\label{eq1}
& \Delta\Phi+\Bigl(\alpha_M+\frac{\alpha_B}{2}\Bigr)\Delta Q = \frac{3}{2}a^2H^2\Omega_m\delta_m \\
\label{eq2}
& \Bigl(\alpha_M+\frac{\alpha_B}{2}\Bigr)\Delta\Phi= \frac{\alpha_B+\alpha_M}{2a^2H^2}\Bigl[(\Delta Q)^2-(\partial_{ij}Q)(\partial_{ij}Q)\Bigr]-\Bigl[\Bigl(\alpha_M+\frac{\alpha_B}{2}\Bigr)^2+\frac{c_s^2}{2}\Bigl(\alpha_K+\frac{3}{2}\alpha_B^2\Bigr)\Bigr] \Delta Q
\end{align}
Notice that for $\alpha_B+\alpha_M=0$, the nonlinearities, and therefore the Vainshtein mechanism, disappear. These models correspond to scalar-tensor theories such as Brans--Dicke, $f(R)$, $\cdots$ . We will work in Fourier space, adopting the following convention
\begin{align}
f(t, \mathbf{r})=\frac{1}{(2 \pi)^3} \int d^3 p f(t, \mathbf{p}) e^{i \mathbf{p} \cdot \mathbf{r}}
\end{align}
The eq.(\ref{eq1},\ref{eq2}) transform into\footnote{We use the same name for the variables $\Phi(k)$, $\Psi(k)$, $\delta \phi(k)$ in the Fourier space.}
\begin{align}
&-{p}^2\Phi(\mathbf{p})-{p}^2\Bigl(\alpha_M+\frac{\alpha_B}{2}\Bigr) Q(\mathbf{p}) = \frac{3}{2}a^2H^2\Omega_m\delta_m(\mathbf{p}) \\
& -{p}^2\Bigl(\alpha_M+\frac{\alpha_B}{2}\Bigr)\Phi(\mathbf{p})
-{p}^2\Bigl[\Bigl(\alpha_M+\frac{\alpha_B}{2}\Bigr)^2+\frac{c_s^2}{2}\Bigl(\alpha_K+\frac{3}{2}\alpha_B^2\Bigr)\Bigr] Q(\mathbf{p})\nonumber\\
&\qquad = \frac{\alpha_B+\alpha_M}{2a^2H^2}\frac{1}{(2\pi)^3}\int {\textnormal d}^3 p_1 {\textnormal d}^3 p_2 \delta_D(\mathbf{p}_{12}-\mathbf{p}) {p}_1^2 {p}_2^2 \gamma(\mathbf{p}_1,\mathbf{p}_2)Q(\mathbf{p}_1)Q(\mathbf{p}_2)
\end{align}
where $\delta_D$ is the Dirac delta, $\mathbf{p}_{12}=\mathbf{p}_{1}+\mathbf{p}_{2}$ and the kernel is
\begin{align}
\label{kernel}
\gamma(\mathbf{p}_1,\mathbf{p}_2)= 1-\frac{(\mathbf{p}_1\cdot \mathbf{p}_2)^2}{p_1^2 p_2^2}
\end{align}
From this, we obtain the nonlinear generalized Poisson equation:
\begin{align}
\label{eq3}
&p^2\Phi(\mathbf{p})+\frac{3}{2}\mu F(\phi) a^2 H^2 \Omega_m\delta_m(\mathbf{p})=\frac{1}{(2\pi)^3}\frac{\Sigma F(\phi)-1}{a^2 H^2 (\alpha_M+\frac{\alpha_B}{2})^2}\int {\textnormal d}^3 p_1 {\textnormal d}^3 p_2 \delta_D(\mathbf{p}_{12}-\mathbf{p}) \nonumber\\
&\gamma(\mathbf{p}_1,\mathbf{p}_2)\Bigl(p_1^2\Phi(\mathbf{p}_1)+\frac{3}{2}a^2H^2\Omega_m\delta_m(\mathbf{p}_1)\Bigr)\Bigl(p_2^2\Phi(\mathbf{p}_2)+\frac{3}{2}a^2H^2\Omega_m\delta_m(\mathbf{p}_2)\Bigr)
\end{align}
We have introduced parameters used in the linear regime such as \cite{Ishak:2019aay}
\begin{align}
\label{mu}
\mu &\equiv \frac{G_{\text {growth}}(a)}{G}=\frac{1}{F}\left[1+\frac{2\left(\alpha_M+\frac{\alpha_B}{2} \right)^2}{c_s^2\left(\alpha_K+\frac{3}{2} \alpha_B^2\right)}\right] \\ 
\label{sigma}
\Sigma &\equiv \frac{G_{\text {lensing}}(a)}{G}=\frac{1}{F}\left[1+\frac{\left(\alpha_M+\frac{\alpha_B}{2} \right)\left(\alpha_M+\alpha_B\right)}{c_s^2\left(\alpha_K+\frac{3}{2} \alpha_B^2\right)}\right]
\end{align}
where $G_{\text {growth}}$ is the effective gravitational coupling which is related to the growth of matter perturbation and $G_{\text {lensing}}$ is the effective gravitational coupling associated with lensing. They are identified from the perturbation equations in the linear regime such that 
\begin{align}
-p^2 \Phi &=4 \pi G \mu a^2 \rho_m \delta_m\\
-p^2(\Phi+\Psi) &=8 \pi G \Sigma a^2 \rho_m \delta_m
\end{align}
In the special case where $\alpha_M+\alpha_B / 2=0$, corresponding to No Slip Gravity \cite{Linder:2018jil}, the effective gravitational couplings for growth and lensing become equal, $G_{\text {growth }}=G_{\text {lensing }}$. Moreover, the interaction at linear order becomes scale-free, \cite{Amendola:2019laa}.

\section{Averaged effective gravitational constant}

In order to introduce the non-linear effective gravitational constant at all scales, we need to introduce the response function of the potential $\Phi$ to a density fluctuation $\delta_m$. In the linear regime, we have from eq.(\ref{eq3})
\begin{align}
    \Phi(\mathbf{p})\propto \frac{1}{p^2} \delta_m(\mathbf{p})
\end{align}
where $1/p^2$ can be interpreted as the linear propagator in the language of renormalized perturbation theory. Formally, the response of the potential to the matter density perturbation can be expanded in the following form \cite{Scoccimarro:2009eu}
\begin{align}
\label{renormalization}
\Phi(t,\mathbf{p})
& =\int {\textnormal d}^3 \mathbf{p}'\left\langle\frac{\mathcal{D} \Phi(t,\boldsymbol{p})}{\mathcal{D} \delta_m(t,\boldsymbol{p}^{\prime})}\right\rangle \delta_m(t,\boldsymbol{p}^{\prime}) +\frac{1}{2!}\int {\textnormal d}^3 \mathbf{p}_1{\textnormal d}^3 \mathbf{p}_2\left\langle\frac{\mathcal{D}^2 \Phi(t,\boldsymbol{p})}{\mathcal{D} \delta_m(t,\boldsymbol{p}_1) \mathcal{D} \delta_m(t,\boldsymbol{p}_2)}\right\rangle \delta_m(t,\boldsymbol{p}_1) \delta_m(t,\boldsymbol{p}_2)\nonumber\\
&+\frac{1}{3!}\int {\textnormal d}^3 \mathbf{p}_1 {\textnormal d}^3 \mathbf{p}_2 {\textnormal d}^3 \mathbf{p}_3\,
\left\langle\frac{\mathcal{D}^3 \Phi(t,\boldsymbol{p})}{\mathcal{D} \delta_m(t,\boldsymbol{p}_1) \mathcal{D} \delta_m(t,\boldsymbol{p}_2) \delta_m(t,\boldsymbol{p}_3)}\right\rangle
\Bigl(
\delta_m(t,\mathbf{p}_1)\delta_m(t,\mathbf{p}_2)\delta_m(t,\mathbf{p}_3)\nonumber\\
& 
 -\langle\delta_m\delta_m\rangle(t,p_1,p_2)\,\delta_m(t,\mathbf{p}_3)
-\text{cyc.}\Bigr)
+\ldots
\end{align}
where $\mathcal{D}$ stands for a functional derivative, while $\delta_m(t,\mathbf{p})$ stands for the \emph{fully non‑linear} density contrast, and \emph{all} functional derivatives are taken with respect to this same field.  
To prevent loop momenta already resummed inside $\delta_m$ from being counted a second time, we retain only the \emph{connected part} of every derivative, in the sense of the cumulant (linked‑cluster) expansion \cite{Scoccimarro:2009eu}.  
The square bracket in the $n=3$ term of Eq.\,(\ref{renormalization}) represents the connected three‑point cumulant of the field~$\delta_m$.  
Without the subtractions proportional to $\langle\delta_m\delta_m\rangle\,\delta_m$ the first product
$\delta_m(\mathbf p_1)\delta_m(\mathbf p_2)\delta_m(\mathbf p_3)$
would, upon Wick contraction, generate diagrams in which two external legs form a
self‑contained ``bubble’’ loop; the loop momentum carried by that bubble has already been integrated over at lower (second) order.
Removing these disconnected pieces therefore guarantees that every loop momentum appears exactly once in the perturbative hierarchy, preserving a one‑to‑one correspondence between diagrams and terms in the functional expansion. Because of translation invariance, we should have
\begin{align}
&\Gamma^{(1)}(t,p) \delta_{\mathrm{D}}(\boldsymbol{p}-\boldsymbol{p}^{\prime}) \equiv\Bigl\langle\frac{\mathcal{D} \Phi(t,\boldsymbol{p})}{\mathcal{D} \delta_m(t,\boldsymbol{p}^{\prime})}\Bigr\rangle\\
&\Gamma^{(2)}\left(t,\boldsymbol{p}_1, \boldsymbol{p}_2\right) \delta_{\mathrm{D}}\left(\boldsymbol{p}-\boldsymbol{p}_{12}\right) \equiv \frac{1}{2!}\left\langle\frac{\mathcal{D}^2 \Phi(t,\boldsymbol{p})}{\mathcal{D} \delta_m(t,\boldsymbol{p}_1) \mathcal{D} \delta_m(t,\boldsymbol{p}_2)}\right\rangle
\end{align}
which simplifies the expression (\ref{renormalization}) 

\begin{align}
\label{eq:PHI}
\Phi(t,\mathbf{p})
=\Gamma^{(1)}(t,p) \delta_m(t,\boldsymbol{p})+\int {\textnormal d}^3{p}_1  {\textnormal d}^3{p}_2 \Gamma^{(2)}\left(t,\boldsymbol{p}_1, \boldsymbol{p}_2\right)\delta_m(t,\boldsymbol{p}_1) \delta_m(t,\boldsymbol{p}_2) \delta_D(\boldsymbol{p}_{12}-\boldsymbol{p})  +\ldots
\end{align}
To understand these expressions, let us look for a perturbative solution
\begin{align}
\Phi(\mathbf{p})=\Phi^{(1)}(\mathbf{p})+\Phi^{(2)}(\mathbf{p})+\Phi^{(3)}(\mathbf{p})+\cdots
\end{align}
where $\Phi^{(n)}$ is proportional to $n$-terms of $\delta_m$. We also expand the two-point propagator itself
\begin{align}
\Gamma^{(1)}(p)=\Gamma_{\text {tree }}^{(1)}(p)+\Gamma_{1 \text {-loop }}^{(1)}(p)+\Gamma_{2 \text {-loop }}^{(1)}(p)+\cdots,
\end{align}
and similarly we can expand the three point propagator and the higher propagators. Perturbatively, we can find
\begin{align}
p^2 \Phi^{(1)}(\mathbf{p}) &= - A_\mu \delta_m(\mathbf{p})\\
p^2 \Phi^{(2)}(\mathbf{p}) &= B (A-A_\mu)^2 \int \gamma(\mathbf{p}_1,\mathbf{p}_2)\delta_m(\mathbf{p}_1) \delta_m(\mathbf{p}_2)\delta_D(\mathbf{p}_{12}-\mathbf{p}) {\textnormal d}^3 p_1 {\textnormal d}^3 p_2
\end{align}
where
\begin{align}
A &= \frac{3}{2} a^2 H^2 \Omega_m\\
A_\mu &= \frac{3}{2}\mu F(\phi) a^2 H^2 \Omega_m\\
B & = \frac{1}{(2\pi)^3}\frac{\Sigma F(\phi)-1}{a^2 H^2 (\alpha_M+\frac{\alpha_B}{2})^2}
\end{align}
From which we can get
\begin{align}
p^2\Bigl\langle\frac{\mathcal{D} \Phi^{(1)}(\boldsymbol{p})}{\mathcal{D} \delta_m(\boldsymbol{p}^{\prime})}\Bigr\rangle = -A_\mu \delta_D(\boldsymbol{p}-\boldsymbol{p}')
\end{align}
and therefore
\begin{align}
\label{eq:tree}
\Gamma^{(1)}_{\text{tree}}(p) = -\frac{A_\mu}{p^2}
\end{align}
Similarly, we get
\begin{align}
\Bigl\langle\frac{\mathcal{D} \Phi^{(2)}(\boldsymbol{p})}{\mathcal{D} \delta_m(\boldsymbol{p}^{\prime})}\Bigr\rangle = 0\,,\quad\text{because  }\langle \delta_m \rangle = 0
\end{align}
Therefore the 1-loop correction has to be obtained from $\Phi^{(3)}$
\begin{align}
p^2\Bigl\langle\frac{\mathcal{D} \Phi^{(3)}(\boldsymbol{p})}{\mathcal{D} \delta_m(\boldsymbol{p}^{\prime})}\Bigr\rangle &= 2 (2\pi)^3 B^2 (A-A_\mu)^3 \delta_D(\boldsymbol{p}-\boldsymbol{p}')\int \gamma(\boldsymbol{p}_1,\boldsymbol{p}_2) \Bigl[\gamma(\boldsymbol{p},-\boldsymbol{p}_1) P(p_1)\nonumber\\
&+\gamma(\boldsymbol{p},-\boldsymbol{p}_2) P(p_2)\Bigr] \delta_D(\boldsymbol{p}_{12}-\boldsymbol{p}){\textnormal d}^3 p_1 {\textnormal d}^3 p_2
\end{align}
where we introduced the power spectrum as follows
\begin{align}
\Bigl\langle \delta_m(\mathbf{p}) \delta_m(\mathbf{k}) \Bigr\rangle &= (2\pi)^3 \delta_D(\mathbf{p}+\mathbf{k}) P(k)
\end{align}
From which we obtain
\begin{align}
\label{eq:1loop}
\Gamma^{(1)}_{\text{1-loop}}(p) = 
\frac{2}{p^2} (2\pi)^3 B^2 (A-A_\mu)^3 \int \gamma(\boldsymbol{p}_1,\boldsymbol{p}_2) \Bigl[\gamma(\boldsymbol{p},-\boldsymbol{p}_1) P(p_1)
+\gamma(\boldsymbol{p},-\boldsymbol{p}_2) P(p_2)\Bigr] \delta_D(\boldsymbol{p}_{12}-\boldsymbol{p}){\textnormal d}^3 p_1 {\textnormal d}^3 p_2
\end{align}
Similarly these calculations can be performed for the three point propagator and the tree level is
\begin{align}
\label{eq:bis-tree}
\Gamma^{(2)}_{\text{tree}}(\boldsymbol{p}_1,\boldsymbol{p}_2) = \frac{B(A-A_\mu)^2}{p^2}\gamma(\boldsymbol{p}_1,\boldsymbol{p}_2)
\end{align}
The tree level expression is dominant at large scales while the loop corrections contribute at smaller scales. The two-point (linear) propagator $\Gamma^{(1)}$ describes how $\Phi$ respond linearly to $\delta_m$. It captures not only the linear theory, but captures the renormalized or dressed linear relation between $\Phi(\mathbf{p})$ and $\delta_m(\mathbf{p})$. If everything stayed linear, the resulting density/ potential fields would be Gaussian if the initial conditions were Gaussian. Because $\Gamma^{(1)}$ is effectively the "linear kernel" (though with loop corrections), it governs how power spectra evolves, but on its own, it does not directly generate non Gaussian features at large scales. On the other side, the three‐point propagator describes how $\Phi(\mathbf{p})$ responds quadratically in $\delta_m$, which implies mode-coupling: large-scale modes can be influenced by products of smaller-scale (or similarly large-scale) density modes. Such mode-coupling is precisely what generates non-Gaussian features, we no longer have a purely linear, one-to-one map from $\delta_m(\mathbf{p})$ to $\Phi(\mathbf{p})$. Instead, $\Phi(\mathbf{p})$ at large scales can pick up contributions from the product of modes $\delta_m\left(\mathbf{p}_1\right) \delta_m\left(\mathbf{p}_2\right)$. It may seem surprising that a second-order kernel affects large scales, since we could associate nonlinearity with small-scale structure formation. But mode-coupling can feed correlations into large scales. Thus $\Gamma^{(2)}$ directly encodes the leading non-Gaussian correction that appears at long wavelengths beyond the linear, Gaussian approximation. In conclusion, the two-point propagator is relevant for the modification of the gravitational constant at all scales and therefore the power spectrum while the three-point propagator is relevant for the bispectrum.

\section{Averaged spherical approximation}

In this section, we will use a spherical approximation, to gain some intuition and also will be a starting point of a later iteration. In the "spherical approximation", the two wave-vectors $\hat{p}_1, \hat{p}_2$ can point in any directions with equal probability, therefore using instead of the kernel (\ref{kernel}), the angular average of the kernel
\begin{align}
    \langle \gamma(\hat{p}_1, \hat{p}_2) \rangle = \int_0^\pi\int_0^{2\pi} (1-\cos^2\theta)\sin\theta d\theta d\phi=\frac{8\pi}{3} = \frac{2}{3} \int \sin\theta d\theta d\phi
\end{align}
we can replace $\gamma$ by $2/3$ in the "spherical approximation", which simplifies  eq.(\ref{eq3}). Finally using an inverse Fourier transform, we find
\begin{align}
-\Delta\Phi+\frac{3}{2} \mu F(\phi) a^2 H^2 \Omega_m \delta_m=\frac{2}{3} \frac{\Sigma F(\phi)-1}{a^2 H^2\left(\alpha_M+\frac{\alpha_B}{2}\right)^2}\Bigl(-\Delta \Phi+\frac{3}{2} a^2 H^2 \Omega_m \delta_m\Bigr)^2
\end{align}
which gives
\begin{align}
\label{eq:Geffspherical}
\Delta\Phi = \frac{3}{2}a^2H^2\Omega_m\delta_m+\frac{9}{8}\frac{a^2H^2\Omega_m(\mu F(\phi)-1)}{g}\Bigl(\sqrt{1+\frac{8}{3}g\delta_m}-1\Bigr)
\end{align}
where we defined the coefficient
\begin{align}
g(a)=\frac{3}{2}\Omega_m\frac{(\Sigma F(\phi)-1)(\mu F(\phi)-1)}{(\alpha_M+\frac{\alpha_B}{2})^2}
\end{align}
At low density contrast, we recover the linear regime with
\begin{align}
\Delta\Phi = \frac{3}{2}a^2H^2\Omega_m\mu F\delta_m
\end{align}
while at high densities, we recover general relativity as expected in the Vainshtein mechanism
\begin{align}
\Delta\Phi = \frac{3}{2}a^2H^2\Omega_m\delta_m
\end{align}
Notice that $F(\phi)$ is not screened\footnote{It could be screened via a chameleon mechanism near massive objects.}, because we have at high densities
\begin{align}
\Delta\Phi = \frac{3}{2}a^2H^2\Omega_m\delta_m = 4\pi \frac{G}{F}a^2\rho_m\delta_m
\end{align}
In our approach, $F(\phi)$ is not screened which might be related to the quasi-static approximation considered in this work. Further analysis of this point might be considered in the future.

Using this spherical approximation (\ref{eq:Geffspherical}), we can roughly estimate the propagators where we have clearly abused notation by considering $\Gamma$ in the Fourier space while $\delta_m$ is in the real space 
\begin{align}
\Gamma^{(1)}(t,p) & \simeq -\frac{1}{p^2}\frac{\partial \Delta \Phi}{\partial \delta_m} = -\frac{1}{p^2}\frac{3}{2}a^2 H^2 \Omega_m \Bigl(1+\frac{\mu F-1}{\sqrt{1+\frac{8}{3}g\delta_m}}\Bigr)\\
\Gamma^{(2)}\left(t,p\right) & \simeq -\frac{1}{p^2}\frac{\partial^2 \Delta \Phi}{\partial \delta_m^2} = \frac{1}{p^2} \frac{g a^2 H^2\Omega_m(\mu F-1)}{\Bigl(1+\frac{8}{3}g\delta_m\Bigr)^{3/2}}
\end{align}
For these expressions to have some meaning, we need to rewrite them in the Fourier space, by considering that on average $\delta_m$ can be replaced by the amplitude of density perturbations 
\begin{align}
\Delta(k) = \sqrt{\frac{k^3 P(k)}{2 \pi^2}}
\end{align}
which gives us a naive spherical approximation
\begin{align}
\label{Gamma1spherical}
\Gamma^{(1)}(t,p) & = -\frac{1}{p^2}\frac{3}{2}a^2 H^2 \Omega_m \Bigl(1+\frac{\mu F-1}{\sqrt{1+\frac{8}{3}g\Delta(k)}}\Bigr)\\
\Gamma^{(2)}\left(t,p\right) & = \frac{1}{p^2} \frac{g a^2 H^2\Omega_m(\mu F-1)}{\Bigl(1+\frac{8}{3}g\Delta(k)\Bigr)^{3/2}}
\end{align}
It is easy to see that $\Gamma^{(1)}$ reproduces the tree level (\ref{eq:tree}) and the 1-loop correction (\ref{eq:1loop}) under some approximations even if some additional terms are introduced such as $\sqrt{P(p)}$ if we expand $\Gamma^{(1)}$ in series of $g$. We see therefore that at small scales where the density is large, we have
\begin{align}
    \frac{\Gamma^{(2)}}{\Gamma^{(1)}}\propto \Delta(k)^{-3/2}\ll 1
\end{align}
while at large scales where the density is small
\begin{align}
    \frac{\Gamma^{(2)}}{\Gamma^{(1)}}\simeq -\frac{2g(\mu F-1)}{3\mu F}
\end{align}
which is of the order of the percent. In the model that we will study later, we will see that $g\simeq 0.2$ and $\mu F\simeq 0.9$. Therefore, higher order corrections will be negligible at small scales while at large scales, we will have $\Gamma^{(n)}/\Gamma^{(1)}\propto g^{n-1}$ which therefore makes these contributions negligible. Therefore, we can neglect higher order terms at all scales and consider only $\Gamma^{(1)}$ in our expansion, for the definition of the effective gravitational constant and the corrections to the power spectrum. But as we have already mentioned, the third point propagator will be necessary for corrections to the bispectrum.

\section{Renormalized effective gravitational constant}

The effective gravitational constant is an important object for correct predictions of the weak lensing and galaxy clustering. It is natural to define the effective gravitational constant from the response of the gravitational potential to the density of matter. When we speak of an effective gravitational constant, we're asking, given a single Fourier mode $\delta_m(\mathbf{p})$ in the matter density, how big is the resulting $\Phi(\mathbf{p})$. That is exactly the linear derivative w.r.t. $\delta_m(\mathbf{k})$. Hence, by definition, an effective gravitational constant for one Fourier mode arises from one functional derivative, and therefore arises from $\Gamma^{(1)}$ and not from higher propagators. It is therefore natural to define the non-linear effective gravitational constant or renormalized gravitational constant $G_{\text{eff}}^{\text{NL}}$ such that $G_{\text{eff}}^{\text{NL}}(t,p) \propto \Gamma^{(1)}(t,p)$
\begin{align}
\label{eq:GeffNL}
\frac{G_{\text{eff}}^{\text{(NL)}}(t,p)}{G}=\frac{1}{\frac{3}{2}a^2H^2\Omega_m F}\frac{\partial (-p^2 \Phi)}{\partial \delta_m}=\frac{-p^2 \Gamma^{(1)}(t,p)}{\frac{3}{2}a^2H^2\Omega_m F}
\end{align}
At tree level (\ref{eq:tree}), we recover the linear gravitational constant
\begin{align}
\frac{G_{\text{eff}}^{\text{(tree)}}(t,p)}{G}=\frac{-p^2 \Gamma^{(1)}_{\text{tree}}(t,p)}{\frac{3}{2}a^2H^2\Omega_m F} = \mu
\end{align}
and for example, the 1-loop correction (\ref{eq:tree},\ref{eq:1loop}) is 
\begin{align}
\frac{G_{\text{eff}}^{\text{1-loop}}(t,p)}{G} &=\frac{-p^2 \Gamma^{(1)}_{\text{tree}}(t,p)-p^2 \Gamma^{(1)}_{\text{1-loop}}(t,p)}{\frac{3}{2}a^2H^2\Omega_m F} \\
&= \mu +g^2\frac{(\mu F-1)}{2F(2\pi)^3} \int \gamma(\boldsymbol{p}_1,\boldsymbol{p}_2) \Bigl[\gamma(\boldsymbol{p},-\boldsymbol{p}_1) P(p_1)
+\gamma(\boldsymbol{p},-\boldsymbol{p}_2) P(p_2)\Bigr] \delta_D(\boldsymbol{p}_{12}-\boldsymbol{p}){\textnormal d}^3 p_1 {\textnormal d}^3 p_2
\label{eq:1loopGeff}
\end{align}
where we see clearly the expansion as powers of the coefficient $g$. The objective of this paper is to find the full resummed expansion of the gravitational constant. Finally, we can see that in the spherical approximation, we have
\begin{align}
G_{\text{eff}}^{\text{(spherical)}}(t,p)=\frac{G}{F}\Bigl(1+\frac{\mu F-1}{\sqrt{1+\frac{8}{3}g\Delta(p)}}\Bigr)
\end{align}
Therefore, we recover that at large scales, we have $G_{\text{eff}}/G=\mu$ while at small scales, we have $G_{\text{eff}}=G/F$ confirming again that $F$ is not screened in the Vaishtein mechanism. 

\section{Master equation}
We saw in the spherical approximation, that $\Gamma^{(1)}$ can be written as (\ref{Gamma1spherical}), which leads us to redefine the response function as
\begin{align}
\label{redefinition}
\Gamma^{(1)}(t,p) &= -\frac{1}{p^2}\frac{3}{2} a^2 H^2 \Omega_m\Bigl(1+(\mu F(\phi)-1) M({t,p})\Bigr)
\end{align}
where $M(t,p)$ is an unknown function. Notice that by construction, we have at large scales $M = 1$ while we would expect that $M=0$ at small scales in order to recover general relativity. This new function, $M(t,p)$, is the main variable of our problem, from which we can derive the nonlinear effective gravitational constant (\ref{eq:GeffNL})
\begin{align}
\label{Geff-M}
G_{\text{eff}}^{\text{(NL)}}(t,p) &=\frac{G}{F}\Bigl(1+(\mu F(\phi)-1) M({t,p})\Bigr)
\end{align}
Going back to our problem, using the redefinition (\ref{redefinition}) we obtain
\begin{align}
\Phi(t,\mathbf{p})=\Gamma^{(1)}(t,p)\delta_m (t,\mathbf{p}) = -\frac{1}{p^2}\frac{3}{2} a^2 H^2 \Omega_m\Bigl(1+(\mu F-1) M({t,p})\Bigr) \delta_m (t,\mathbf{p})
\end{align}
from which we derive the identities
\begin{align}
&p^2\Phi+\frac{3}{2}a^2H^2\Omega_m \delta_m = -\frac{3}{2}a^2 H^2\Omega_m (\mu F-1)M\delta_m\\
&p^2\Phi+\frac{3}{2}a^2H^2\Omega_m \mu F \delta_m = -\frac{3}{2}a^2 H^2\Omega_m (\mu F-1)(M-1)\delta_m
\end{align}
Replacing these expressions into the nonlinear Poisson equation~(\ref{eq3}), we obtain 
\begin{align}
\delta_m(\mathbf{p})(M({p})-1) &=-\frac{g}{(2\pi)^3}\int {\textnormal d}^3 p_1 {\textnormal d}^3 p_2 \delta_D(\mathbf{p}_{12}-\mathbf{p})  \gamma(\mathbf{p}_1,\mathbf{p}_2)\delta_m(\mathbf{p}_1)\delta_m(\mathbf{p}_2)M({p}_1)M({p}_2)
\end{align}
where the time dependence is implicit and not written explicitly. Multiplying this equation by $\delta_m(\mathbf{k})$ and taking expectation values, we obtain
\begin{align}
\label{intermediate}
(M({p})-1)P(p)\delta_D(\mathbf{p}+\mathbf{k}) &=-\frac{g}{(2\pi)^3}\int {\textnormal d}^3 p_1 {\textnormal d}^3 p_2 \delta_D(\mathbf{p}_{12}-\mathbf{p})\delta_D(\mathbf{p}_{12}+\mathbf{k})  \gamma(\mathbf{p}_1,\mathbf{p}_2)B({p}_1,{p}_2,{k})M({p}_1)M({p}_2)
\end{align}
where we introduced the bispectrum as follows
\begin{align}
\Bigl\langle \delta_m(\mathbf{p}_1) \delta_m(\mathbf{p}_2)\delta_m(\mathbf{p}_3) \Bigr\rangle &= (2\pi)^3 \delta_D(\mathbf{p}_1+\mathbf{p}_2+\mathbf{p}_3) B({p}_1,{p}_2,p_3)
\end{align}
Eq.(\ref{intermediate}) reduces to
\begin{align}
\label{mastereq-previous}
(M({p})-1)P(p) &=-\frac{g}{(2\pi)^3}\int {\textnormal d}^3 p_1 {\textnormal d}^3 p_2 \delta_D(\mathbf{p}_{12}-\mathbf{p}) \gamma(\mathbf{p}_1,\mathbf{p}_2) B(\mathbf{p}_1,\mathbf{p}_2)M({p}_1)M({p}_2)
\end{align}
which we can easily integrate wrt to $p_2$ and the azimuth to obtain
\begin{align}
\label{mastereq}
P(p)(1-M({p})) &=\frac{g}{(2\pi)^2}\int p^2 \gamma(\mathbf{p}_1,\mathbf{p}-\mathbf{p}_1) B(\mathbf{p}_1,\mathbf{p}-\mathbf{p}_1)M({p}_1)M\Bigl(\sqrt{p^2+p_1^2-2p p_1 \mu_\theta}\Bigr) {\textnormal d} p_1 {\textnormal d} \mu_\theta
\end{align}
where
\begin{align}
\gamma(\mathbf{p}_1,\mathbf{p}-\mathbf{p}_1) = p_1^2\frac{1-\mu_\theta^2}{p^2+p_1^2-2p p_1 \mu_\theta}\,,~~~\text{and}~~ \mathbf{p}\cdot \mathbf{p}_1 = p p_1 \mu_\theta
\end{align}
Given a specific model, with a power spectrum and bispectrum, we could in principle solve  eq.(\ref{mastereq}) where the unknown is the function $M(p)$.

\section{Renormalized power spectrum and bispectrum}

As we have seen in the previous section, starting from a given power spectrum and bispectrum, we can renormalize the effective gravitational constant and obtain the averaged effective gravitational constant at all scales. In this section, we will show that it can be used to renormalize the power spectrum and bispectrum, which can then be used in the first step to correct the gravitational constant. This process can be iterated until convergence, at which point we would obtain the power spectrum and bispectrum, as well as the effective gravitational constant. To do this, we begin with the fluid equations, assuming that matter is nonrelativistic (as usual, we will not consider higher-order corrections such as viscosity)
\begin{align}
& \dot{\delta}_m+\frac{1}{a} \nabla_i\left[(1+\delta_m) u^i\right] =0, \\
& \dot{u}^i+H u^i+\frac{1}{a} u^j \nabla_j u^i =-\frac{1}{a} \nabla^i \Phi .
\end{align}
where the Newtonian potential is given by (\ref{eq:PHI}). These equations can be written in the Fourier space, using the divergence of the velocity $\theta=\nabla_i u^i / a H$ 
\begin{align}
\label{renorm1}
& \frac{\dot{\delta}_m(t, \boldsymbol{p})}{H}+\theta(t, \boldsymbol{p})=-\frac{1}{(2 \pi)^3} \int d^3 p_1 d^3 p_2 \delta_{\text{D}}(\boldsymbol{p}_{12}-\boldsymbol{p}) \alpha(\boldsymbol{p}_1, \boldsymbol{p}_2) \theta(t, \boldsymbol{p}_1) \delta_m(t, \boldsymbol{p}_2) \equiv N_\alpha \\
\label{renorm2}
& \frac{\dot{\theta}(t, \boldsymbol{p})}{H}+\left(2+\frac{\dot{H}}{H^2}\right) \theta(t, \boldsymbol{p})-\frac{p^2}{a^2 H^2} \Phi(t, \boldsymbol{p}) =-\frac{1}{(2 \pi)^3} \int d^3 p_1 d^3 p_2 \delta_{\text{D}}(\boldsymbol{p}_{12}-\boldsymbol{p})  \beta(\boldsymbol{p}_1, \boldsymbol{p}_2) \theta(t, \boldsymbol{p}_1) \theta(t, \boldsymbol{p}_2) \equiv N_\beta\\
\label{renorm3}
& \Phi(t,\mathbf{p})
=\Gamma^{(1)}(t,p) \delta_m(t,\boldsymbol{p})+\int {\textnormal d}^3{p}_1  {\textnormal d}^3{p}_2\Gamma^{(2)}\left(t,\boldsymbol{p}_1, \boldsymbol{p}_2\right)\delta_m(t,\boldsymbol{p}_1) \delta_m(t,\boldsymbol{p}_2)\delta_{\text{D}}(\mathbf{p}_{12}-\mathbf{p})  +\ldots
\end{align}
where we used the kernels
\begin{align}
& \alpha(\boldsymbol{p}_1, \boldsymbol{p}_2)=1+\frac{\left(\boldsymbol{p}_1 \cdot \boldsymbol{p}_2\right)}{p_2^2} \\ 
& \beta(\boldsymbol{p}_1, \boldsymbol{p}_2)=\frac{\left(\boldsymbol{p}_1 \cdot \boldsymbol{p}_2\right)\left|\boldsymbol{p}_1+\boldsymbol{p}_2\right|^2}{2 p_1^2 p_2^2}
\end{align}
The equations (\ref{renorm1}, \ref{renorm2}) can be solved perturbatively to correct the assumed power spectrum and bispectrum. The algorithm is the following

\begin{enumerate}
\item[-] Step 1: Start with an initial power spectrum $P(p)$ and bispectrum $B\left(p_1, p_2, p_3\right)$.
\item[-] Step 2: Use these to solve eq. (\ref{mastereq}) for $M(p)$ and thereby obtain a modified effective gravitational constant.
\item[-] Step 3: Insert this back into the fluid equations (\ref{renorm1},\ref{renorm2},\ref{renorm3}) to obtain corrected $P(p)$ and $B\left(p_1, p_2, p_3\right)$.
\item[-] Step 4: Iterate until convergence.
\end{enumerate}
In this paper, we will focus only on the construction of the effective gravitational constant which will be used in steps 3 and 4 in a future work. But let us develop the formalism for these last steps. We can rewrite this system of equations as
\begin{align}
\delta_m''+\Bigl(2+\frac{\dot H}{H^2}\Bigr) \delta_m' &-\frac{3}{2}\Omega_m F \frac{G_{\text{eff}}^{\text{(NL)}}}{G}\delta_m = N_\alpha'+\Bigl(2+\frac{\dot H}{H^2}\Bigr)N_\alpha-N_\beta\nonumber\\
&-\frac{p^2}{a^2 H^2} \int {\textnormal d}^3{p}_1  {\textnormal d}^3{p}_2\Gamma^{(2)}\left(t,\boldsymbol{p}_1, \boldsymbol{p}_2\right)\delta_m(t,\boldsymbol{p}_1) \delta_m(t,\boldsymbol{p}_2)\delta_{\text{D}}(\mathbf{p}_{12}-\mathbf{p})
\end{align}
where $'\equiv {\textnormal d}/{\textnormal d} \ln a$, and we have neglected higher order propagators as explained in previous sections. The left hand side of this equation is the renormalized part which corrects the power spectrum, while the right hand side would correct the bispectrum, namely the mixing of scales. Therefore, it is enough to solve 
\begin{align}
\label{eq:drenorm}
\delta_m''+\Bigl(2+\frac{\dot H}{H^2}\Bigr) \delta_m' &-\frac{3}{2}\Omega_m F \frac{G_{\text{eff}}^{\text{NL}}}{G}\delta_m = 0
\end{align}
to obtain corrections to the power spectrum. The solution can be written as 
\begin{align}
\label{renorm_deltam}
\delta_m ^{(1)}(t,p) &= D_\text{renorm}(t,p) \delta_1(p)
\end{align}
where $\delta_1(p)$ correspond to the linear initial perturbation while $D_\text{renorm}$ corresponds the renormalized growth factor and not only the linear version in standard perturbation theory because of the nonlinear effective gravitational constant included in this equation. 

For the bispectrum, we can obtain easily the corrections at large scales by using the tree-level propagator (\ref{eq:bis-tree}). Using standard perturbation theory \cite{Bernardeau:2001qr} we can always write 
\begin{align}
\delta_m ^{(1)}(t,p) &= D_{\text{renorm}}(t,p) \delta_1(p) \\
\delta_m ^{(2)}(t,p) &= \frac{1}{(2 \pi)^3}\int {\textnormal d}^3 p_1 {\textnormal d}^3 p_2 \delta_D(\mathbf{p}_{12}-\mathbf{p})  F_2(t,\mathbf{p}_1,\mathbf{p}_2)\delta_1(p_1) \delta_1(p_2)
\end{align}
and from our equations (\ref{renorm1},\ref{renorm2},\ref{renorm3}), we get
\begin{align}
F_2''+\Bigl(2+\frac{\dot H}{H^2}\Bigr) F_2'-\frac{3}{2} \Omega_m F \frac{G_{\text{eff}}^{\text{NL}}}{G}F_2 &=\Bigl(\frac{3}{2} \Omega_m F \frac{G_{\text{eff}}^{\text{NL}}}{G} D_\text{renorm}^2+D_\text{renorm}'^2\Bigr)\alpha(\mathbf{p}_1,\mathbf{p}_2)\nonumber\\
&+ \beta(\mathbf{p}_1,\mathbf{p}_2)D_\text{renorm}'^2-\frac{3}{2}g\Omega_m(\mu F-1)D_\text{renorm}^2 \gamma(\mathbf{p}_1,\mathbf{p}_2)
\end{align}
which we can rewrite using the Legendre polynomials

\begin{align}
&F_2''+\Bigl(2+\frac{\dot H}{H^2}\Bigr) F_2'-\frac{3}{2} \Omega_m F \frac{G_{\text{eff}}^{\text{NL}}}{G}F_2 =
\Bigl(\frac{3}{2} \Omega_m F \frac{G_{\text{eff}}^{\text{NL}}}{G} D_\text{renorm}^2+\frac{4}{3}D_\text{renorm}'^2-g\Omega_m (\mu F-1)D_\text{renorm}^2\Bigr)\nonumber\\
&\qquad +\Bigl(\frac{3}{2} \Omega_m F \frac{G_{\text{eff}}^{\text{NL}}}{G} D_\text{renorm}^2+2D_{\text{renorm}}'^2\Bigr)\frac{1}{2}\Bigl(\frac{p_1}{p_2}+\frac{p_2}{p_1}\Bigr)L_1(\hat{p}_1 \cdot \hat{p}_2)\nonumber\\
&\qquad +\Bigl(\frac{2}{3}D_\text{renorm}'^2+g\Omega_m(\mu F-1)D_\text{renorm}^2\Bigr)
L_2(\hat{p}_1 \cdot \hat{p}_2)
\end{align}
where we defined $\hat{p}=\mathbf{p}/p$. It is therefore obvious that we can write
\begin{align}
F_2(t,\mathbf{p}_1,\mathbf{p}_2) = D_0(t,p)+\frac{1}{2} D_1(t,p)\left(\frac{p_1}{p_2}+\frac{p_2}{p_1}\right) \left(\hat{p}_1 \cdot \hat{p}_2\right)+D_2(t,p) \left(\hat{p}_1 \cdot \hat{p}_2\right)^2
\end{align}
where $p$ is understood as the external momentum ${p}=|\mathbf{p}_1+\mathbf{p}_2|$, with the equations
\begin{align}
\label{eq:D0}
& D_0''+\Bigl(2+\frac{\dot H}{H^2}\Bigr) D_0'-\frac{3}{2} \Omega_m F \frac{G_{\text{eff}}^{\text{NL}}}{G}D_0 =
\frac{3}{2} \Omega_m F \frac{G_{\text{eff}}^{\text{NL}}}{G} D_\text{renorm}^2+\frac{4}{3}D_\text{renorm}'^2-g\Omega_m (\mu F-1)D_\text{renorm}^2 \\
\label{eq:D2}
& D_2''+\Bigl(2+\frac{\dot H}{H^2}\Bigr) D_2'-\frac{3}{2} \Omega_m F \frac{G_{\text{eff}}^{\text{NL}}}{G}D_2 = \frac{2}{3}D_\text{renorm}'^2+g\Omega_m(\mu F-1)D_\text{renorm}^2
\end{align}
and $D_1=D_0+D_2$. The equation for $D_1$ is therefore 
\begin{align}
D_1''+\Bigl(2+\frac{\dot H}{H^2}\Bigr) D_1'-\frac{3}{2} \Omega_m F \frac{G_{\text{eff}}^{\text{NL}}}{G}D_1 =
\frac{3}{2} \Omega_m F \frac{G_{\text{eff}}^{\text{NL}}}{G} D_\text{renorm}^2+2 D_\text{renorm}'^2
\end{align}
In the squeezed limit $p_1\ll p_2$, the only piece that generates the soft pole
$\propto(p_2/p_1)\,\mu_\theta$ with $\mu_\theta=\hat{\mathbf p}_1\!\cdot\!\hat{\mathbf p}_2$
is the $D_1$ term, giving
\begin{align}
F_2^{\text {soft }}(p_1, p_2)=\frac{p_2}{2 p_1} \mu_\theta D_1(t, k)+\text { finite }
\end{align}
In Horndeski with universal (Jordan-frame) matter coupling, the fluid+gravity equations enjoy the same extended Galilean invariance as in $\Lambda$CDM. That symmetry fixes the soft limit of the kernels: no new soft poles appear, and the universal “shift” piece (the $\propto \mathbf p_2\!\cdot\!\mathbf p_1/p_1^2$ term when $\mathbf p_1\to0$) is the same as in GR. What changes is only the finite tidal part of the squeezed limit, through the (time/scale-dependent) linear growth~\cite{Crisostomi:2019vhj}. The Ward identity then fixes the soft piece \cite{Peloso:2013spa}:
\begin{align}
F_2^{\mathrm{soft}}(p_1, p_2)=\frac{p_2}{2 p_1} \mu_\theta\left[D_{\text {renorm }}(t, k)\right]^2+\text { finite. }
\end{align}
so comparing the two expressions enforces
\begin{align}
D_1(t, k)=\left[D_{\text {renorm }}(t, k)\right]^2\,.
\end{align}
This result can be easily recovered. Indeed, if we define the differential operator
\begin{align}
    \mathcal{L}[X]=X''+\Bigl(2+\frac{\dot H}{H^2}\Bigr) X'-\frac{3}{2} \Omega_m F \frac{G_{\text{eff}}^{\text{NL}}}{G}X
\end{align}
we have $\mathcal{L}[D_{\text{renorm}}]=0$ but more interestingly
\begin{align}
\mathcal{L}\Bigl[D_{\text{renorm}}^2\Bigr] = \frac{3}{2} \Omega_m F \frac{G_{\text{eff}}^{\text{NL}}}{G} D_\text{renorm}^2+2 D_\text{renorm}'^2
\end{align}
hence
\begin{align}
    \mathcal{L}[D_1]=\mathcal{L}\Bigl[D_{\text{renorm}}^2\Bigr]
\end{align}
from which we conclude
\begin{align}
    D_1(t,p) = c(p) D_{\text{renorm}}(t,p)+D_{\text{renorm}}^2(t,p)
\end{align}
where $c(p)$ is an integration constant. 

For Gaussian adiabatic initial conditions (IC), at the initial epoch the connected bispectrum must vanish, so $\delta^{(2)}$ must start at order $D^2$. Therefore the IC are
\begin{align}
D_1\left(t_i, p\right)=0, \quad D_1^{\prime}\left(t_i, p\right)=0,
\end{align}
which kills the homogeneous piece and gives
\begin{align}
D_1=D_{\text {renorm }}^2 .
\end{align}
Therefore the second-order kernel satisfies the EP/extended-Galilean soft-limit constraint; deviations from GR appear only in the finite (monopole/quadrupole) coefficients through the modified linear growth.

In conclusion, given a renormalized effective gravitational constant, obtained at previous steps, we can in principle evolve easily eqs.(\ref{eq:drenorm},\ref{eq:D0},\ref{eq:D2}) and consequently obtain the corrections to the power spectrum and the bispectrum which will be used in eq.(\ref{mastereq}) to correct the gravitational constant. These additional steps will be performed in a future work while we will address the construction of the renormalized gravitational constant in the next sections.

\section{The model}

\begin{figure}[H]
    \centering
    \includegraphics[width=0.45\linewidth]{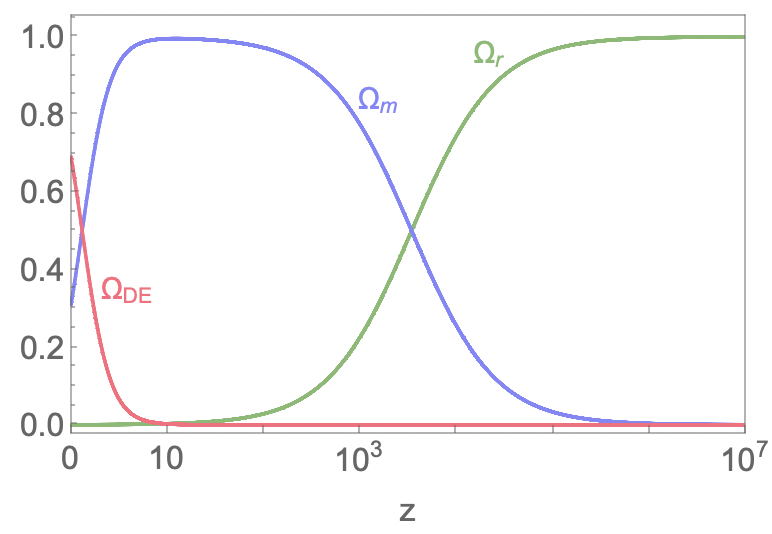}
    \includegraphics[width=0.48\linewidth]{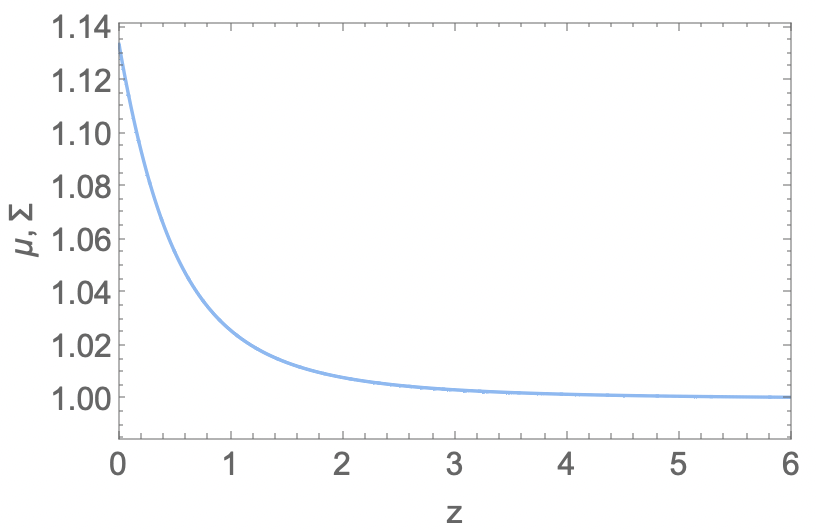}
    \caption{On the left, the evolution of the density parameters $\Omega_m, \Omega_r$, and $\Omega_{DE}$ as a function of the redshift $z$, while on the right, the evolution of the linear effective gravitational constant, $\mu=\Sigma$, as a function of the redshift $z$.}
    \label{fig:1}
\end{figure}

\begin{figure}[H]
    \centering
    \includegraphics[width=0.49\linewidth]{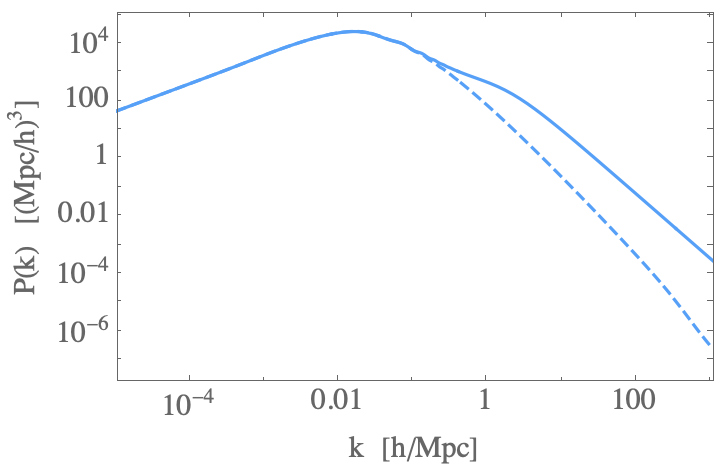}
    \includegraphics[width=0.49\linewidth]{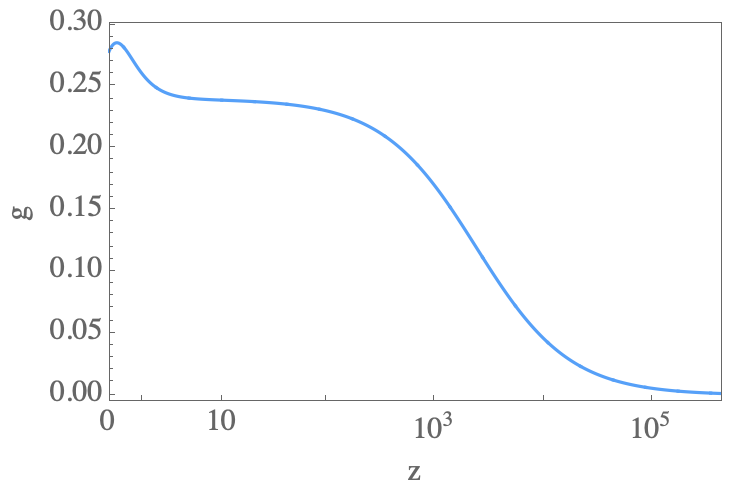}
    \caption{On the left, the power spectrum as a function of the scale $k$ in units of $h / \mathrm{Mpc}$, where the dashed line represents the linear power spectrum and the solid line corresponds to the nonlinear Halofit correction. On the right, the evolution of the parameter $g$ as a function of the redshift $z$.}
    \label{fig:2}
\end{figure}

To proceed with the analysis, we must adopt a specific model. As noted earlier, the coupling function $F(\phi)$ is not screened by the Vainshtein mechanism. To avoid fine-tuning, we assume $F(\phi)=1$. Rather than selecting a particular Lagrangian by specifying the remaining functions $K(\phi,X)$ and $G(\phi,X)$, it is more practical to use a parametrization for the functions $(\alpha_K, \alpha_B, \alpha_M)$. Since we have already imposed $\alpha_T=0$, and our choice of $F(\phi)=1$ leads to $\alpha_M=0$, this parametrization allows us to separate the background evolution from the perturbation dynamics. Therefore, we adopt a conservative approach for the background assuming a standard $\Lambda$CDM evolution while the perturbations are described using the following parametrization
\cite{Bellini:2014fua}
\begin{align}
    \alpha_K = c_K \Omega_{DE}\,,\quad \alpha_B = c_B \Omega_{DE}.
\end{align}
For our analysis, we take $c_K=1$ and $c_B=0.5$ and evolve the system using \texttt{hi\_class} \cite{Blas:2011rf,Zumalacarregui:2016pph,Bellini:2019syt}. As we have mentioned the background evolution follows the standard $\Lambda$CDM model with cosmological parameters taken from \cite{Planck:2018vyg}. Figure \ref{fig:1} shows the evolution of the density parameters and the linear gravitational constants associated with matter growth (\ref{mu}) and lensing (\ref{sigma}).
Figure \ref{fig:2} shows the power spectrum of this model in the linear regime, computed using \texttt{hi\_class}, and its nonlinear extension obtained via the \texttt{halofit} approach \cite{Smith:2002dz}. The same figure also shows the function $g(z)$, which plays a crucial role in our analysis. The perturbative framework requires this coefficient to remain below unity for all redshifts, with its present-day value given in our model by $g\simeq 0.279$

\section{Numerical analysis}

In the previous section, we have constructed the basic elements required to solve (\ref{mastereq}). Starting with an initial guess for \(M(p)\), the right-hand side of (\ref{mastereq}) can be integrated to yield a new function \(M(p)\) until reaching convergence. We have considered four different approaches to solve this problem that we will develop in the next sections because that justifies our final approach which differs from \cite{Scoccimarro:2009eu}. For all our attempts, our initial guess of \(M(p)\) is derived from the spherical approximation (\ref{Gamma1spherical}) which gives us
\begin{align}
M^{(\text{spherical})}(p) = \frac{1}{\sqrt{1+\frac{8}{3}g\Delta(p)}}
\end{align}
The power spectrum is obtained at a given redshift from \texttt{hi\_class} while for the bispectrum we will use two different approaches. In the approach 1 (A1) we will obtain the bispectrum from the code \texttt{BiHalofit} \cite{Takahashi:2019hth} while, in a second approach, we will assume a simpler approach (A2) by using the tree-level bispectrum from standard perturbation theory. Its form is
\begin{align}
\mathrm{B}_{\text{tree}}\left({p}_1, {p}_2, {p}_3\right)=2 F_2\left({p}_1, {p}_2,{p}_3\right) {P}(p_1) {P}({p}_2)+\text{ cyclic permutations, }
\end{align}
where \(F_2\) is the kernel
\begin{align}
F_2\left({p}_1, {p}_2, {p}_3\right)=\frac{5}{7}+\frac{\mu_\theta}{2}\Bigl(\frac{p_1}{p_2}+\frac{p_2}{p_1}\Bigr) +\frac{2}{7} \mu_\theta^2 \quad \text{with} \quad \mu_\theta=\frac{p_3^2-p_1^2-p_2^2}{2p_1 p_2}.
\end{align}
While for the power spectrum used to construct the bispectrum, we consider the non-linear version in order to capture some non-linearities.

The reader interested in the best numerical approach can go directly to the section (\ref{method4}) instead of studying the failed approaches.

\subsection{Method 1}
\label{method1}

\begin{figure}[H]
    \centering
    \includegraphics[width=0.48\linewidth]{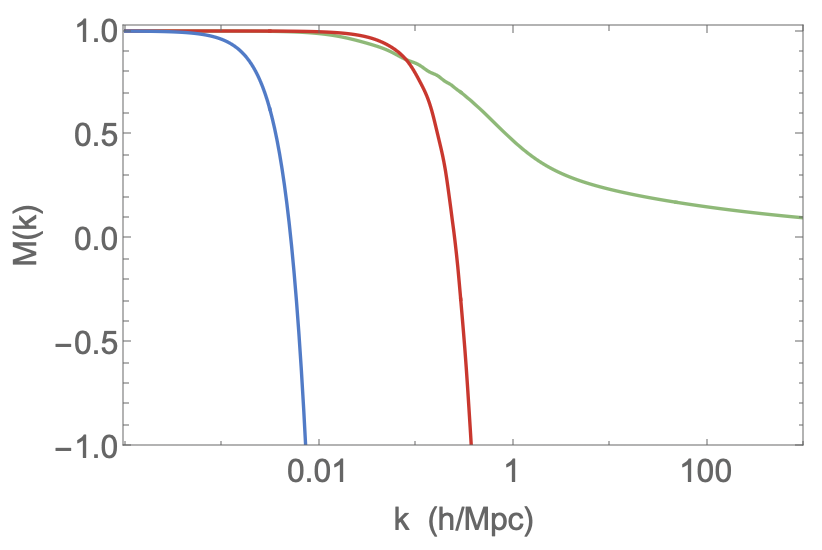}
    \includegraphics[width=0.48\linewidth]{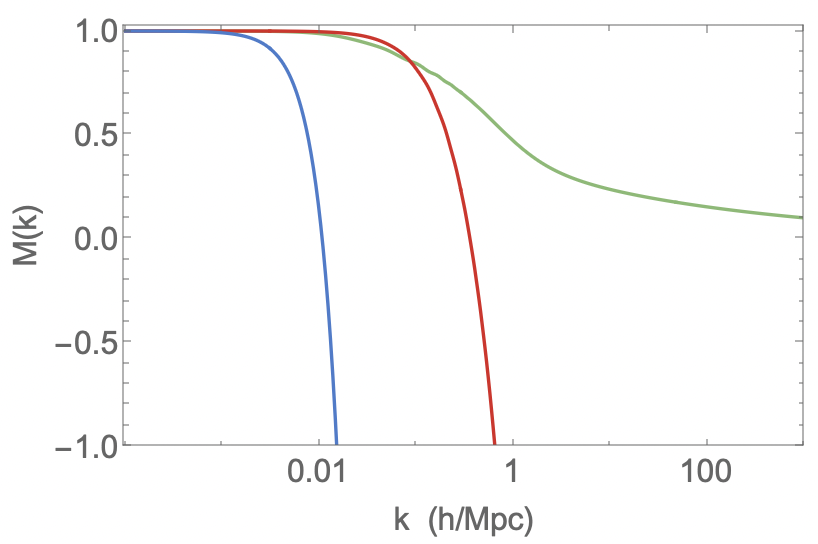}
    \caption{We show the iterations on $M(p)$ at $z=0$ for the A1 bispectrum on the left and A2 bispectrum on the right. The green, red, and blue curves correspond to the initial spherical guess $M^{(0)}$, the first iteration $M^{(1)}$, and the second iteration, respectively.}
    \label{fig:method1}
\end{figure}

In this section, we intend to solve directly eq.(\ref{mastereq}) by assuming our two approaches for the bispectrum, A1 and A2. The integrals are performed using \texttt{CUBA} \cite{Hahn:2004fe} in particular the routine \texttt{Vegas} but other routines have been also used as a consistency check. The results are shown in Figure (\ref{fig:method1}). We see that no convergence is achieved. The function $M(p)$ is more negative at each iteration. If we increase the precision in the integral, $M(p)$ continues to go negative but less rapidly. We conclude that the system is extremely sensitive to the precision of the integral and only an infinite precision could generate the right convergence. This approach is therefore not viable and that explains why we adopt the other approaches.

\subsection{Method 2}

In this section, we follow the approach used in \cite{Scoccimarro:2009eu} in the analysis of the DGP model. In order to avoid the problems presented in the previous section, we multiply the main equation (\ref{mastereq-previous}) by \(e^{i\mathbf{p}\cdot\mathbf{r}}\) and integrating over $p$, we obtain
\begin{align}
\xi(r)-\chi(r)=\frac{g}{(2\pi)^6}\int {\textnormal d}^3 p_1 {\textnormal d}^3 p_2 e^{i\mathbf{p}_{12}\cdot \mathbf{r}}\gamma(\mathbf{p}_1,\mathbf{p}_2) B(\mathbf{p}_1,\mathbf{p}_2)M({p}_1)M({p}_2)
\end{align}
where we defined the correlation functions
\begin{align}
\xi(r) &=\frac{1}{(2\pi)^3}\int {\textnormal d}^3\mathbf{p}\, e^{i \mathbf{p} \cdot \mathbf{r}} P(p)\\
\chi(r) &=\frac{1}{(2\pi)^3}\int {\textnormal d}^3\mathbf{p}\, e^{i \mathbf{p} \cdot \mathbf{r}} P(p) M(p)
\end{align}
For numerical stability, it is advantageous \cite{Scoccimarro:2009eu} to define a new function \(Q_{\text{eff}}\) such that\footnote{The introduction of \(Q_{\text{eff}}\) in the eq.(\ref{eq:Qeff}) is dictated by the structure of the integral. The nonlinear term appears through the convolution of two factors of $M(p)$ with the bispectrum kernel. When we Fourier transform this convolution to configuration space, it naturally produces a term proportional to $\chi^2(r)$.}
\begin{align}
\label{eq:Qeff}
Q_{\text{eff}}(r) \chi^2(r) = \frac{1}{(2\pi)^6} \int {\textnormal d}^3 p_1 {\textnormal d}^3 p_2 e^{i\mathbf{p}_{12}\cdot \mathbf{r}} \gamma(\mathbf{p}_1,\mathbf{p}_2) B(\mathbf{p}_1,\mathbf{p}_2) M(p_1) M(p_2),
\end{align}
which transforms eq.(\ref{mastereq}) into  
\begin{align}
\xi(r) - \chi(r) = g Q_{\text{eff}}(r) \chi^2(r),
\end{align}
leading to the solution  
\begin{align}
\chi(r) = \frac{\sqrt{1+4 g Q_{\text{eff}}(r) \xi(r)}-1}{2 g Q_{\text{eff}}(r)}.
\end{align}
Here, we retain only the positive root, consistent with the linear regime. After applying an inverse Fourier transform, we obtain
\begin{align}
\label{eq:M}
M(p) = \frac{2\pi}{P(p)} \int_0^{\infty} r^2 J_0(pr) \frac{\sqrt{1+4 g Q_{\text{eff}}(r) \xi(r)}-1}{g Q_{\text{eff}}(r)} \,{\textnormal d}r,
\end{align}
where \(J_0\) is the zeroth spherical Bessel function or sinc function. In summary, the calculation proceeds in three main steps:
\begin{enumerate}
    \item Starting with an initial \(M(p)\), compute \(\xi(r)\) and \(\chi(r)\) using for example \texttt{FFTLog} \cite{Fang:2019xat}.
    \item Evaluate \(Q_{\text{eff}}(r)\) from eq. (\ref{eq:Qeff}).
    \item Calculate the updated \(M(p)\) using eq. (\ref{eq:M}) via another \texttt{FFTLog}.
\end{enumerate}
After completing step 3, the new \(M(p)\) is used to restart from step 1, iterating this procedure until \(M(p)\) converges. The primary computational effort lies in evaluating the integral in eq.(\ref{eq:Qeff}) which we can reduce to a 3D integral. For that we perform first a change of variables, passing from an integral over $(p_1,p_2)$ to an integral over $(p_1,p_3)$ using $\mathbf{p}_1+\mathbf{p}_2+\mathbf{p}_3=\mathbf{0}$
\begin{align}
I(r) &= \int {\textnormal d}^3 p_1 {\textnormal d}^3 p_2 e^{i\mathbf{p}_{12}\cdot \mathbf{r}}\gamma(\mathbf{p}_1,\mathbf{p}_2) B(\mathbf{p}_1,\mathbf{p}_2)M({p}_1)M({p}_2)\\[1mm]
&= \int {\textnormal d}^3 p_1 {\textnormal d}^3 p_3 e^{-i\mathbf{p}_{3}\cdot \mathbf{r}}\gamma(\mathbf{p}_1,-\mathbf{p}_1-\mathbf{p}_3) B(p_1,p_2,p_3)M({p}_1)M({p}_2)\,,\quad p_2=\sqrt{p_1^2+p_3^2+2\mathbf{p}_1\cdot \mathbf{p}_3}
\end{align}
We can always define $\mathbf{r}$ in the $z$ direction which gives us $\mathbf{p}_3\cdot \mathbf{r}=p_3 r \cos\theta_3$ with $\theta_3$ the angle between $\mathbf{p}_3$ and the $z$ direction. Also $\mathbf{p}_1$ is defined in the basis where $\mathbf{p}_3$ is the z direction which gives us $\mathbf{p}_1\cdot \mathbf{p}_3=p_1p_3\cos\theta_{13}$ where $\theta_{13}$ is the angle between $\mathbf{p}_3$ and $\mathbf{p}_1$. The integrals on the azimuth angles are trivial, giving
\begin{align}
I(r) &= (2\pi)^2 \int {\textnormal d} p_1 {\textnormal d} p_3 {\textnormal d} \mu_3 {\textnormal d} \mu_{13} \, p_1^2 p_3^2  e^{-ip_3 r \mu_3}\gamma(\mathbf{p}_1,-\mathbf{p}_1-\mathbf{p}_3) B(p_1,p_2,p_3)M({p}_1)M({p}_2)
\end{align}
where $\mu_3=\cos\theta_3$ and $\mu_{13}=\cos\theta_{13}$. The integral over $\mu_3$ is also trivial and we obtain
\begin{align}
\label{eq:Ir3D}
I(r) &= 2 (2\pi)^2 \int {\textnormal d} p_1 {\textnormal d} p_3 {\textnormal d} \mu_{13} \, p_1^2 p_3^2  J_0(p_3 r)\gamma(\mathbf{p}_1,-\mathbf{p}_1-\mathbf{p}_3) B(p_1,p_2,p_3)M({p}_1)M({p}_2)
\end{align}
where
\begin{align}
p_2 = \sqrt{p_1^2+p_3^2+2p_1p_3\mu_{13}}\,,\quad\text{and}~~ 
\gamma(\mathbf{p}_1,-\mathbf{p}_1-\mathbf{p}_3) = \frac{p_3^2(1-\mu_{13}^2)}{p_1^2+p_3^2+2p_1p_3\mu_{13}}.
\end{align}
This integral can be evaluated directly using \texttt{CUBA} or we can notice that the \(r\) dependence can be factorized out and therefore for each value of $p_{3}$ we can evaluate the 2 dimensional integral
\begin{align}
\label{eq:Fp3}
F(p_3)=2 (2\pi)^2 \int {\textnormal d} p_1 {\textnormal d} \mu_{13} \, p_1^2 \gamma(\mathbf{p}_1,-\mathbf{p}_1-\mathbf{p}_3) B(p_1,p_2,p_3)M({p}_1)M({p}_2)
\end{align}
and finally an \texttt{FFTLog} over \(p_3\) gives the final result
\begin{align}
\label{eq:Ir}
I(r) = \int {\textnormal d} p_3 \, p_3^2 F(p_3) J_0(p_3 r)
\end{align}
The Bessel function causes oscillations in $M(p)$. However, when we split the integration into two steps, first calculating the 2D integral \ref{eq:Fp3} and then applying an \texttt{FFTLog} over $p_3$ \ref{eq:Ir} gives a much better result. In contrast, directly evaluating the 3D integral \ref{eq:Ir3D} without using \texttt{FFTLog} produces large oscillations, especially at larger scales. 

\begin{figure}[H]
\centering
\includegraphics[width=0.48\linewidth]{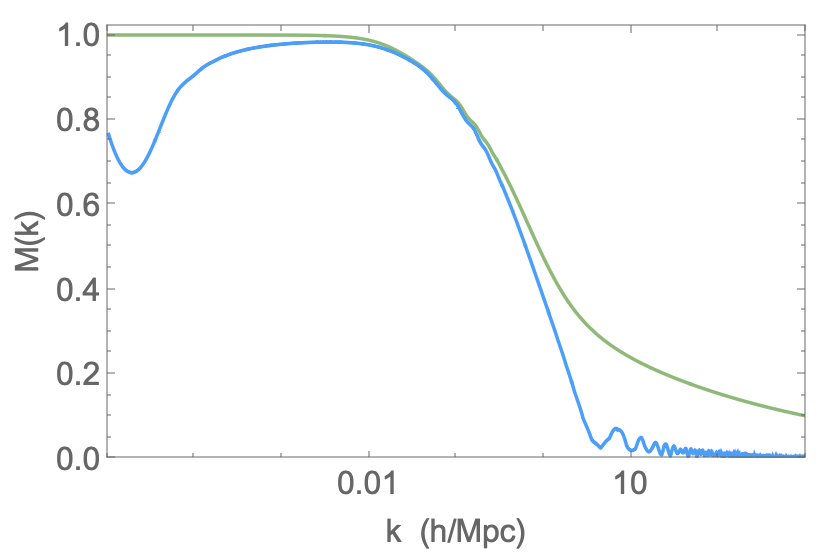}
\includegraphics[width=0.48\linewidth]{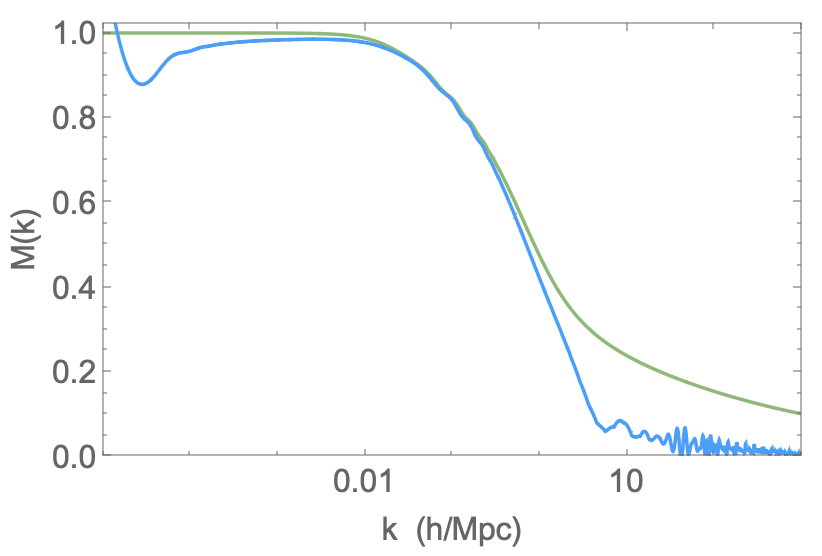}
\caption{We show the first iteration on $M(p)$ at $z=0$ for the A1 bispectrum on the left and A2 bispectrum on the right. The green and blue curves correspond to the initial spherical guess $M^{(0)}$ and the first iteration $M^{(1)}$ respectively.}
\label{fig:method2}
\end{figure}

All methods, however, exhibit some oscillatory behavior, and in order to reduce them, we have chosen to compute these integrals for small scales, $r$, and rely on our initial guess for large scales. The outcomes for both bispectra, A1 and A2, are shown in Fig.\ref{fig:method2}. After just one iteration, there is a significant deviation at large scales, where no change is expected, and the oscillations at small scales become increasingly amplified with each iteration. While it would be possible to smooth $M(p)$ for the subsequent steps, doing so would be overly artificial.

\subsection{Method 3}

In this method, we provide an alternative way to calculate the integral in eq.(\ref{eq:Qeff}), which will be performed using a \texttt{2D-FFTLog} approach \cite{Fang:2020vhc}. To achieve this, we reduce the integral in eq. (\ref{eq:Qeff}) to  
\begin{align}
\label{eq:qeff2}
Q_{\text{eff}}(r)\chi^2(r) = \frac{1}{4 \pi^4} \sum_{\ell=0}^\infty (-1)^\ell \iint j_\ell(p_1 r) j_\ell(p_2 r) M(p_1) M(p_2) B_\ell(p_1, p_2) p_1^2 p_2^2 {\textnormal d}p_1 {\textnormal d}p_2,
\end{align}
where \(j_\ell(x)\) are the spherical Bessel functions, and  
\begin{align}
B_\ell(p_1, p_2) = \frac{2\ell+1}{2} \int_{-1}^1 (1-x^2) B(p_1, p_2, p_3) P_\ell(x) {\textnormal d}x, \quad 
p_3 = \sqrt{p_1^2 + p_2^2 + 2p_1 p_2  x},
\end{align}
with \(P_\ell(x)\) being the Legendre polynomials. Further details are provided in the appendix \ref{Appendix-2DFFTLog}.

\begin{figure}[H]
\centering
\includegraphics[width=0.48\linewidth]{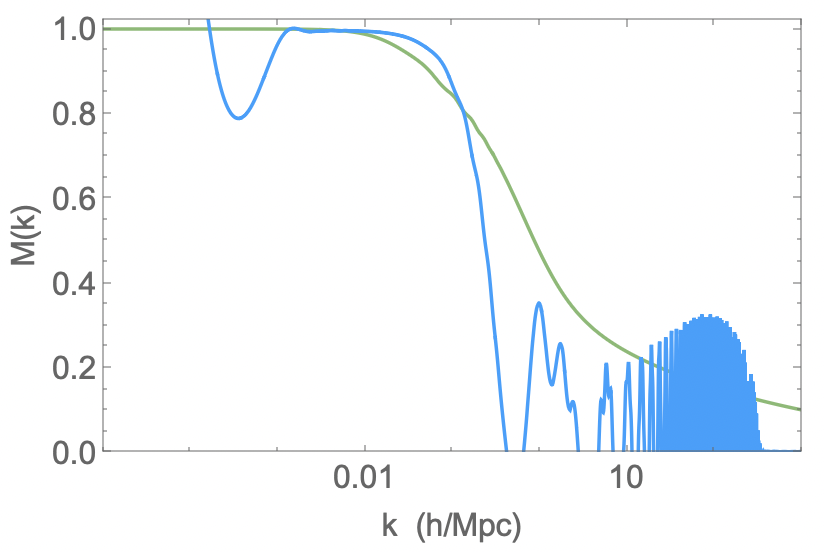}
\includegraphics[width=0.48\linewidth]{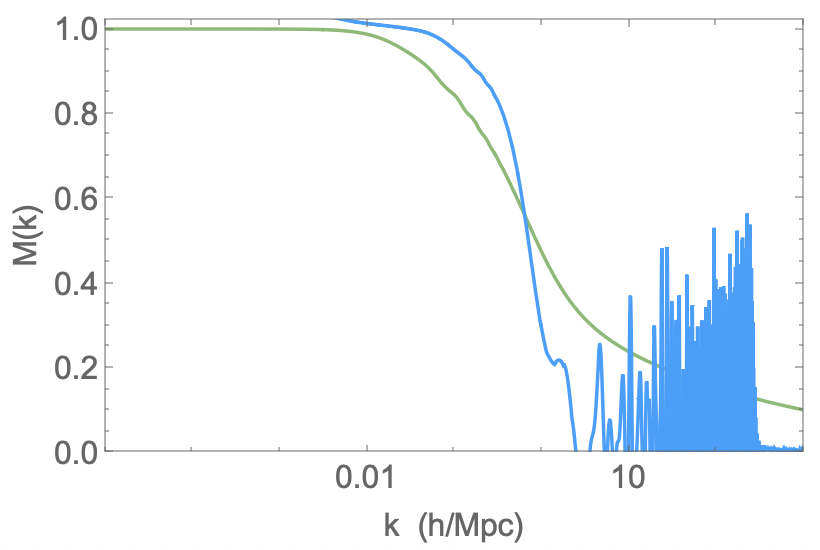}
\caption{We show the first iteration on $M(p)$ at $z=0$ for the A1 bispectrum on the left and A2 bispectrum on the right. The green and blue curves correspond to the initial spherical guess $M^{(0)}$ and the first iteration $M^{(1)}$ respectively.}
\label{fig:method3}
\end{figure}

Using the initial bispectrum, the integral above can be evaluated, for example, via a trapezoidal method for different values of \(\ell\). We can easily verify that we have achieved a good order in maximal value of \(\ell\) if the inverse transformation reproduces with good precision our original bispectrum
\begin{align}
(1-x^2) B\Bigl(p_1,p_2,\sqrt{p_1^2+p_2^2+2p_1p_2x}\Bigr) = \sum_{\ell=0}^{\ell_{\text{max}}} B_\ell(p_1,p_2) P_\ell(x).
\end{align}
We found that good results were achieved with $\ell_{\text{max}}=6$. Once the $B_\ell$ functions are obtained, equation $(\ref{eq:qeff2})$ can be integrated using \texttt{2D-FFTLog}. Note that at each iteration, only the \texttt{2D-FFTLog} needs to be evaluated because only $M(p)$ changes, while the $B_\ell$ functions remain fixed, because independent of $M(p)$. As shown in Fig.(\ref{fig:method3}), significant oscillations appear, primarily due to the \texttt{2D-FFTLog} and the precision in calculating of $B_\ell$. Importantly, we have not restricted the evaluation of these integrals to small scales and retained large-scale data from our initial guess as did in the method 2. In this approach, $M(p)$ is evaluated at all scales.

\subsection{Method 4}
\label{method4}

\begin{figure}[H]
\centering
\includegraphics[width=0.48\linewidth]{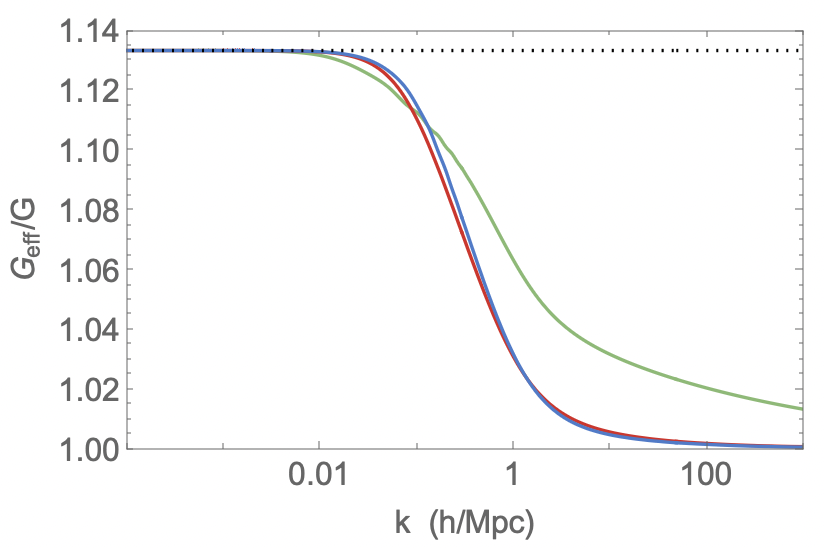}
\includegraphics[width=0.48\linewidth]{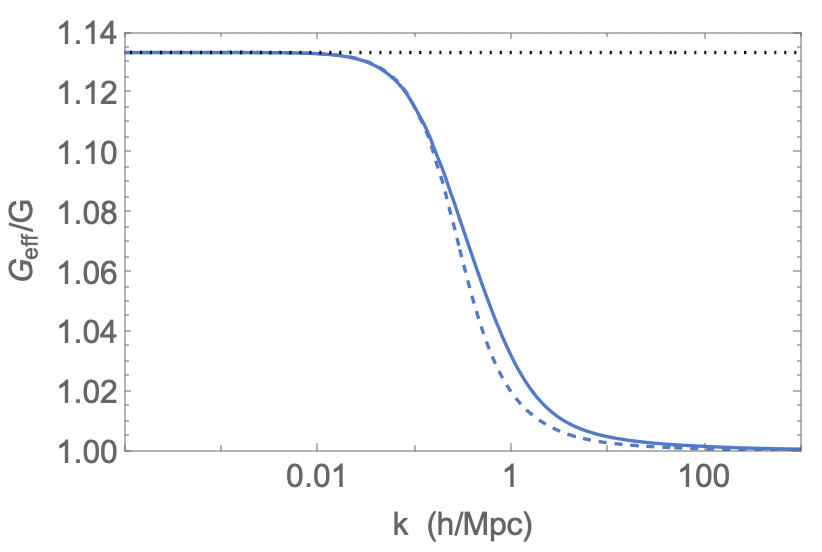}
\caption{On the left, we show the convergence of the renormalized gravitational constant at $z=0$. The dotted line indicates the linear gravitational constant, while the green, red, and blue curves correspond to the initial spherical guess $M^{(0)}$, the first iteration $M^{(1)}$, and the second iteration, respectively. For this analysis, the A2 bispectrum was employed. On the right, the dashed line represents the converged renormalized gravitational constant for the A1 bispectrum, and the thick line shows the corresponding result for the A2 bispectrum.}
\label{fig:convergence}
\end{figure}

Finally, we show in this section the method we have used. For that, we will consider a modified version of the method 1. Let us define the right hand side of eq.(\ref{mastereq}) as
\begin{align}
\operatorname{RHS}[M]=\frac{g}{(2\pi)^2}\int p^2 \gamma(\mathbf{p}_1,\mathbf{p}-\mathbf{p}_1) B(\mathbf{p}_1,\mathbf{p}-\mathbf{p}_1)M({p}_1)M\Bigl(\sqrt{p^2+p_1^2-2p p_1 \mu_\theta}\Bigr) {\textnormal d} p_1 {\textnormal d} \mu_\theta.
\end{align}
Therefore the naive fixed‐point iteration of section(\ref{method1}) can be rewritten as
\begin{align}
M^{(n+1)}(k)=1-\frac{\operatorname{RHS}\left[M^{(n)}\right]}{P(k)},
\end{align}
which goes negative as we have seen. The right hand side is proportional to \(M(p)^2\) and therefore it is natural to define a new function
\begin{align}
F^{(n)}(k)=\frac{\operatorname{RHS}[M^{(n)}]}{\left(M^{(n)}(k)\right)^2},
\end{align}
which converts our main equation (\ref{mastereq}) into 
\begin{align}
P(k)\left[1-M^{(n+1)}(k)\right]=F^{(n)}(k)\left(M^{(n+1)}(k)\right)^2,
\end{align}
giving
\begin{align}
M^{(n+1)}(k)=\frac{-P(k)\pm \sqrt{P(k)^2+4 P(k) F^{(n)}(k)}}{2 F^{(n)}(k)}.
\end{align}
By construction, the negative root is negative and will evolve as the solution obtained iteratively in the method 1. On the contrary, the positive root remains positive and bounded from above by \(1\). This approach transforms the iteration into a quadratic relationship that enforces positivity. In conclusion, we solve for a given \(M(p)\) the RHS of eq.(\ref{mastereq}) which gives us \(F(k)\) once divided by \(M(p)^2\). We therefore obtain the new \(M(p)\) by the formula
\begin{align}
\frac{-P(k)+ \sqrt{P(k)^2+4 P(k) F(k)}}{2 F(k)}.
\end{align}
We found with this method a convergence with only 2 iterations. Once \(M(p)\) is obtained we have the renormalized gravitational constant using eq.(\ref{Geff-M}). The iteration is seen in Figure (\ref{fig:convergence}). We see that convergence is rapidly obtained without oscillations. Also, the form of $G_{\text{eff}}^{\text{NL}}$ is independent of the bispectrum until $k=0.1$. The slight difference at smaller scale should disappear once corrections to the bispectrum and principally to the power spectrum are taken into account. 

\begin{figure}[H]
\centering
\includegraphics[width=0.48\linewidth]{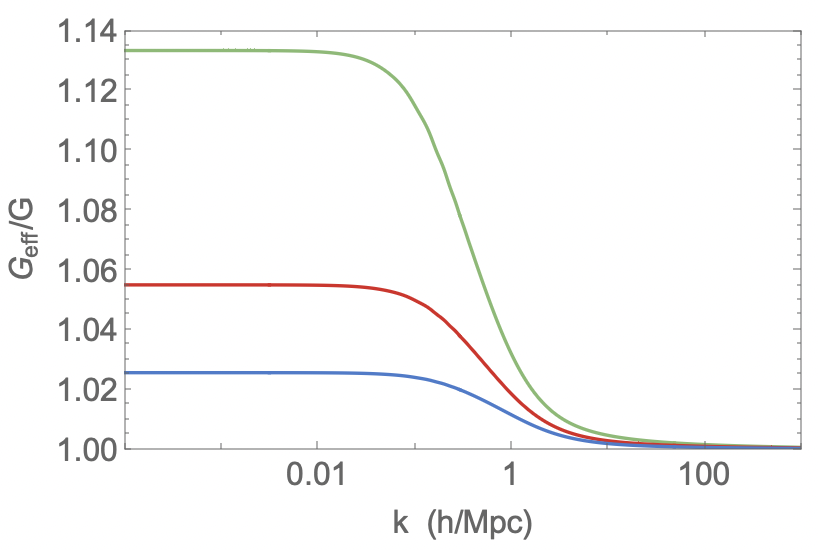}
\includegraphics[width=0.48\linewidth]{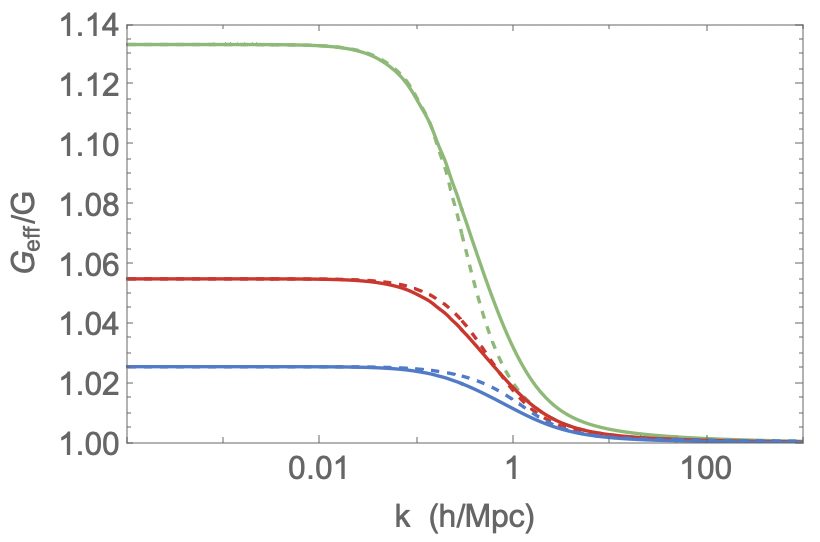}
\caption{On the left, we show the converged renormalized gravitational constant at $z=0$, $0.5$ and $1$, in green, red and blue respectively for the A2 bispectrum. On the right, the dashed lines represent the A1 bispectrum and the thick lines correspond to the A2 bispectrum.}
\label{fig:Geff-z}
\end{figure}

Using this method, we can calculate the renormalized power spectrum at various redshifts and thereby determine $G_{\text{eff}}^{\text{NL}}(k,z)$ which is necessary for the evaluation of the corrections to the power spectrum (see Eq. (\ref{renorm_deltam})). In Fig.\ref{fig:Geff-z}, we show the renormalized gravitational constant for $z=0$, $z=0.5$ and $z=1$. If we define the Vainshtein radius as the scale below which $G_{\text{eff}}^{\text{NL}}$ is reduced by, say, $90\%$, then the Vainshtein radius was smaller in the past. However, if we instead define the Vainshtein radius as the scale at which the gravitational constant attains a specific value, let us say $G_{\text{eff}}^{\text{NL}}=1.01 G$, then this radius increases with increasing redshift.

\section{Comparison with Standard Perturbation Theory at 1-Loop}
\begin{figure}[H]
\centering
\includegraphics[width=0.48\linewidth]{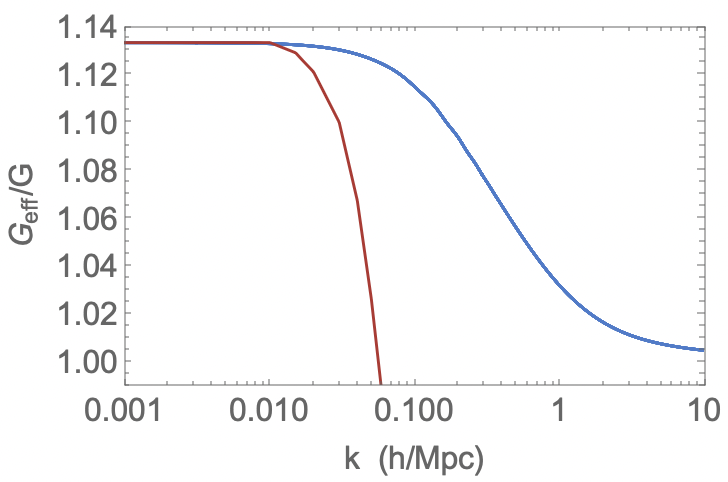}
\caption{We show the gravitational constant at $z=0$. The blue line indicates the renormalized gravitational constant, while the red line corresponds to the SPT approach.}
\label{fig:SPT}
\end{figure}

In the previous sections, we developed the formalism of the renormalized approach to the nonlinear effective gravitational constant, which reduces to GR at small scales. This formalism allows us to identify the scale at which the Vainshtein mechanism becomes effective. To assess the advantages of this approach, we now compare it with the standard perturbation theory (SPT) by performing the same calculations within the SPT framework.\\
In SPT, all perturbation fields, $\Phi$, $\theta$, $\delta_m$, and $Q$, are expanded as
\begin{align}
Y(t, \mathbf{p})=\sum_{n=1} Y_n(t, \mathbf{p})
\end{align}
where $Y$ stands for any of the perturbation fields, and $Y_n$ represents the $n$th-order term in the expansion. At linear order, one finds
\begin{align}
\delta_m^{(1)}(t, \mathbf{p})=D_{+}(t) \delta_{1}(\mathbf{p})\,,\qquad -p^2 \Phi_1(t, \mathbf{p})=\frac{3}{2}a^2 H^2 \Omega_m \mu F D_{+}(t)  \delta_{\mathrm{L}}(\mathbf{p}),
\end{align}
where $D_+(t)$ is the growing mode of the linear evolution equation (\ref{eq:drenorm})
\begin{align}
D''+\Bigl(2+\frac{\dot H}{H^2}\Bigr) D' -\frac{3}{2}\Omega_m F \mu D = 0
\end{align}
with primes denoting derivatives with respect to $\ln a$. The second-order solution $(\delta_m^{(2)}(t, \mathbf{p}),\Phi_{2}(t, \mathbf{p}))$ has been derived in \cite{Takushima:2013foa}, and the third-order $(\delta_m^{(3)}(t, \mathbf{p}),\Phi_{3}(t, \mathbf{p}))$ in \cite{Takushima:2015iha}. Knowing this expansion we can obtain the effective gravitational constant. Indeed, from Eq.(\ref{eq:GeffNL}), we have 
\begin{align}
    -p^2 \Phi(t,\mathbf{p}) = \frac{3}{2}a^2 H^2 \Omega_m F \frac{G_{\text{eff}}^{\text{NL}}}{G} \delta_m(t,\mathbf{p})
\end{align}
contracting both sides with $\delta_m(t,-\mathbf{p})$ and taking an ensemble average, we obtain
\begin{align}
    -p^2 P_{\Phi\delta} = \frac{3}{2}a^2 H^2 \Omega_m F \frac{G_{\text{eff}}^{\text{NL}}}{G} P_{\delta\delta}
\end{align}
from which we get
\begin{align}
\label{GeffSPT}
    \frac{G_{\text{eff}}^{\text{NL}}(t,{p})}{G}  = -\frac{p^2}{\frac{3}{2}a^2 H^2 \Omega_m F} \frac{P_{\Phi\delta}(t,{p})}{P_{\delta\delta}(t,{p})} 
\end{align}
where we have defined the cross correlation function
\begin{align}
\Bigl\langle \Phi(\mathbf{p}) \delta_m(\mathbf{k}) \Bigr\rangle &= (2\pi)^3 \delta_D(\mathbf{p}+\mathbf{k}) P_{\Phi\delta}(p)
\end{align}
and $P_{\delta\delta}\equiv P(k)$. Details of the calculations are provided in Appendix~\ref{SPT1loop}. At 1-loop order, we obtain from Eq.~(\ref{GeffSPT}) the following expression
\begin{align}
    \frac{G_{\text{eff}}^{\text{NL}}}{G}  & \simeq -\frac{p^2}{\frac{3}{2}a^2 H^2 \Omega_m F} \left(\frac{P_{\Phi\delta}^{\text{lin.}}}{P_L}+\frac{P_{\Phi\delta}^{\text{1-loop}}}{P_L}-\frac{P_{\Phi\delta}^{\text{lin.}}}{P_L}\frac{P_{\delta\delta}^{\text{1-loop}}}{P_L}\right) \\
    & = \mu -\frac{p^2}{\frac{3}{2}a^2 H^2 \Omega_m F} \frac{P_{\Phi\delta}^{\text{1-loop}}}{P_L}-\mu\frac{P_{\delta\delta}^{\text{1-loop}}}{P_L}
\end{align}
where we defined
\begin{align}
    \langle \delta_1(t,\mathbf{p}) \delta_1(t,\mathbf{k})\rangle = (2\pi)^3 \delta_D(\mathbf{p}+\mathbf{k})P_L(p)
\end{align}
The numerical results are shown in Fig.~(\ref{fig:SPT}), which illustrates that the 1-loop correction does not fully capture the behavior of the gravitational constant in the non-linear regime.
At very large scales and early times (the IR limit), the 1-loop power is controlled by the hard region, where the loop momentum $q$ is finite so that $p \rightarrow 0$ implies $r=q / p \rightarrow \infty$. The soft region $q \ll p$ is subleading. One finds
\begin{align}
    P_{\delta\delta}^{\text{1-loop}}(p\rightarrow 0) &=2P_{\delta\delta}^{13}(p\rightarrow 0)+\mathcal{O}(p^4)\\
    &= \frac{147-144\lambda_1-64\lambda_2}{630\pi^2}p^2P_L(p)\int_0^\infty dq P_L(q)+\mathcal{O}(p^4)
\end{align}
where $\lambda_1$ and $\lambda_2$ are defined in Eqs. (\ref{eq:lambda1},\ref{eq:lambda2}). Likewise,
\begin{align}
    P_{\Phi\delta}^{\text{1-loop}}(p\rightarrow 0) &=P_{\Phi\delta}^{13}(p\rightarrow 0)+P_{\Phi\delta}^{31}(p\rightarrow 0)+\mathcal{O}(p^2)\\
    &=-\frac{a^2 H^2}{\pi^2} P_L(p)\left[\kappa_\Phi\left(\frac{7}{30}-\frac{8}{35}\lambda_1-\frac{32}{315}\lambda_2\right)+\frac{8}{15}\left(\sigma_\Phi+\mu_\Phi\right)\right]\int_0^\infty dq P_L(q) +\mathcal{O}(p^2)
\end{align}
with $(\sigma_\Phi,\mu_\Phi)$ given in Eqs. (\ref{sigmaphi},\ref{muphi}). Hence, in the IR,
\begin{align}
\label{eq:Geffmatter}
    \frac{G_{\text{eff}}^{\text{NL}}}{G}
    & = \mu +\left[\frac{1}{\Omega_m F} \left(\kappa_\Phi\left(\frac{7}{45}-\frac{16}{105}\lambda_1-\frac{64}{945}\lambda_2\right)+\frac{16}{45}\left(\sigma_\Phi+\mu_\Phi\right)\right)\right.\nonumber\\
    &\qquad \qquad \qquad \qquad \left.-\mu\frac{147-144\lambda_1-64\lambda_2}{630}\right]\frac{p^2}{\pi^2}\int_0^\infty dq P_L(q)+\mathcal{O}(p^4)
\end{align}
We can check that this matches the expected GR matter-era limit: taking
$\mu=\Omega_m=F=\lambda_1=\lambda_2=1$, $\kappa_\Phi=3/2$ and $\sigma_\Phi=\mu_\Phi=0$, all coefficients cancel and
\begin{align}
    \frac{G_{\text{eff}}^{\text{NL}}}{G} = 1+\mathcal{O}(p^4)
\end{align}
In summary, at very large scales in the early Universe the one-loop correction scales as \(p^2\). Assuming matter domination, the coefficients in \eqref{eq:Geffmatter} follow from the time dependences of the growing and decaying modes, which during the matter era read \cite{Calderon:2019vog,Gannouji:2008jr}
\begin{align}
    D_\pm(t)=t^{p_\pm}\,,\quad p_\pm = \frac{-1\pm \sqrt{1+24 F \mu}}{6}
\end{align}
From these relations one obtains \(\lambda_1, \lambda_2, \kappa_\Phi, \sigma_\Phi\) and \(\mu_\Phi\), so that \eqref{eq:Geffmatter} becomes
\begin{align}
    \frac{G_{\text{eff}}^{\text{NL}}}{G} = \mu -A g \frac{\mu F-1}{F} p^2+\mathcal{O}(p^4)
\end{align}
with
\begin{align}
    A = \frac{32}{15 \pi^2} \frac{X(X-1)+2 g(6+X-X^2)}{(X-1)(3X-1)}\int_0^\infty dq P_L(q)
\end{align}
where $X=\sqrt{1+24\mu F}$.\\
We now verify that the renormalized approach yields the same IR behavior, as it should in the linear regime. In the IR limit (the squeezed bispectrum), the leading term is
\begin{align}
    B(\mathbf{p}_1,\mathbf{p}-\mathbf{p}_1)=P_L(p)\left(\Bigl(\frac{13}{7}+\frac{8}{7}\mu_\theta^2\Bigr)P_L(p_1)-p_1 \mu_\theta^2 P_L'(p_1)\right)+\mathcal{O}(p)
\end{align}
Using \eqref{mastereq}, the leading-order behavior of \(M(p)\) is
\begin{align}
    M(p)=1-\frac{g}{(2\pi)^2}p^2 \int dq P_L(q) M(q)^2 \left(\frac{292}{105}-\frac{4}{15}\frac{d\ln P_L}{d\ln q}\right)+\mathcal{O}(p^3)
\end{align}
This equation is solved recursively until convergence. Once converged, the integral is a number, so we write
\begin{align}
    \frac{1}{(2\pi)^2}\int dq P_L(q) M(q)^2 \left(\frac{292}{105}-\frac{4}{15}\frac{d\ln P_L}{d\ln q}\right) = B  
\end{align}
which implies 
\begin{align}
    M(p)=1-B g p^2 +\mathcal{O}(p^3)
\end{align}
Remark that in the IR regime the source in the master equation is non-local: $\sim p^2 \int d q P_L(q)$ (after factoring the external $P_L(p)$), not $\propto P_L(k)^2$. Hence a naive small-$P$ expansion $M \simeq 1- c P+\ldots$ does not apply. From \eqref{Geff-M}, the renormalized effective gravitational constant in the IR limit is
\begin{align}
    \frac{G_{\text{eff}}^{\text{NL}}}{G} = \mu -B g \frac{\mu F-1}{F} p^2+\mathcal{O}(p^3)
\end{align}
The IR limits coincide if $B=A$ namely
\begin{align}
    x\int dq P_L(q) M(q)^2 \left(\frac{292}{105}-\frac{4}{15}\frac{d\ln P_L}{d\ln q}\right) = \frac{128}{15} \frac{X(X-1)+2 g(6+X-X^2)}{(X-1)(3X-1)}\int_0^\infty dq P_L(q)
\end{align}
Even if the IR limit cannot be checked analytically, we conclude that the \(p^2\) scaling is recovered in both approaches. At the next order a difference may already appear: the renormalized approach scales as \(p^3\) (if absent cancellations), whereas the one-loop correction scales as \(p^4\).

\section{Conclusions}

In this paper, we have developed the formalism to obtain the renormalized gravitational constant at all scales for a class of Horndeski models, as well as the renormalized power spectrum and the bispectrum at large scales necessary for evaluating the contributions to non-Gaussianities. The theory is conceptually divided into two parts: the effective gravitational constant, which controls the Vainshtein mechanism and can be treated at first order, and the second-order perturbations that contribute to non-Gaussianities through scale coupling. Interestingly, this renormalized approach to the perturbations depends on powers of a parameter $g$, with $g<1$, which justifies the suppression of higher-order propagators.

We derived the master equation for the nonlinear effective gravitational constant, which means solving a nonlinear integral equation. We have tested various numerical methods and found that the method 4 is the fastest and most precise. Using this method, we obtained the renormalized gravitational constant for a toy model and showed that it is weakly dependent on the form of the bispectrum. Consequently, the next step is to refine the power spectrum and bispectrum to achieve convergence on the gravitational constant. We have also compared our result to a straightforward 1-loop calculation using SPT. However, this latter approach fails to fully capture the correct behavior of $G_{\text{eff}}^{\text{NL}}$ at small scales. But both approaches agree in the IR limit.

Finally, in future work we plan to obtain $G_{\text{eff}}^{\text{NL}}$ at different redshifts, which will serve as the starting point for the renormalization of the power spectrum and bispectrum. A second iteration of this process will then correct the renormalized gravitational constant until convergence is achieved.

\section*{Acknowledgments}

This work is supported by ANID FONDECYT Regular No. 1220965 (Chile). R.G. acknowledges interesting discussions with Jorge Noreña. LA acknowledges support by the Deutsche Forschungsgemeinschaft (DFG, German Research Foundation) under Germany's Excellence Strategy EXC 2181/1 - 390900948 (the Heidelberg STRUCTURES Excellence Cluster) and under Project  554679582 "GeoGrav: Cosmological Geometry
and Gravity with non-linear physics". We are grateful to the anonymous referee for the careful reading and valuable suggestions that helped improve the paper.

\section*{Appendix}
\appendix

\section{Integral used in the 2D-\texttt{FFTLog}}
\label{Appendix-2DFFTLog}

We want to reduce the integral (\ref{eq:Qeff})
\begin{align}
\int {\textnormal d}^3 p_1 {\textnormal d}^3 p_2 e^{i\mathbf{p}_{12}\cdot \mathbf{r}} \gamma(\mathbf{p}_1,\mathbf{p}_2) B(\mathbf{p}_1,\mathbf{p}_2) M(p_1) M(p_2)
\end{align}
For that we will use the following decomposition
\begin{align}
e^{i\mathbf{k}\cdot \mathbf{r}}=4\pi \sum_{\ell=0}^\infty \sum_{m=-\ell}^{\ell} i^\ell j_\ell(k r)  Y_{\ell m}(\hat{\mathbf{k}}) Y_{\ell m}^{*}(\hat{\mathbf{r}})
\end{align}
which gives
\begin{align}
&\int {\textnormal d}^3 p_1 {\textnormal d}^3 p_2 e^{i\mathbf{p}_{12}\cdot \mathbf{r}} \gamma(\mathbf{p}_1,\mathbf{p}_2) B(\mathbf{p}_1,\mathbf{p}_2) M(p_1) M(p_2)=(4\pi)^2\sum_{\substack{\ell_1,m_1\\ \ell_2,m_2}}i^{\ell_1+\ell_2} \int {\textnormal d}^3 p_1 {\textnormal d}^3 p_2  \gamma(\mathbf{p}_1,\mathbf{p}_2) \nonumber \\
&\quad B(\mathbf{p}_1,\mathbf{p}_2) M(p_1) M(p_2) j_{\ell_1}(p_1 r) j_{\ell_2}(p_2 r)Y_{\ell_1 m_1}(\hat{\mathbf{p}}_1)Y_{\ell_2 m_2}^*(\hat{\mathbf{p}}_2)Y_{\ell_1 m_1}^*(\hat{\mathbf{r}})Y_{\ell_2 m_2}(\hat{\mathbf{r}})
\end{align}
The left hand side of this integral is independent of the orientation of $\mathbf{r}$ which means that the integral is a function of the norm $r$, let us call this integral $F(r)$. Integrating over the angles in the space of $\mathbf{r}$ gives trivially
\begin{align}
\iint F(r) d\Omega = 4\pi F(r)
\end{align}
Integrating also the right hand side wrt to the angles associated to the $r$-space, we can reduce the expression using the normalization
\begin{align}
\int Y_{\ell_1 m_1}^*(\hat{\mathbf{r}})Y_{\ell_2 m_2}(\hat{\mathbf{r}}) d\Omega = \delta_{\ell_1\ell_2}\delta_{m_1 m_2}
\end{align}
\begin{align}
&\int {\textnormal d}^3 p_1 {\textnormal d}^3 p_2 e^{i\mathbf{p}_{12}\cdot \mathbf{r}} \gamma(\mathbf{p}_1,\mathbf{p}_2) B(\mathbf{p}_1,\mathbf{p}_2) M(p_1) M(p_2)=4\pi\sum_{\ell,m}(-1)^{\ell} \int {\textnormal d}^3 p_1 {\textnormal d}^3 p_2  \gamma(\mathbf{p}_1,\mathbf{p}_2) \nonumber \\
&\qquad \qquad B(\mathbf{p}_1,\mathbf{p}_2) M(p_1) M(p_2) j_{\ell}(p_1 r) j_{\ell}(p_2 r)Y_{\ell m}(\hat{\mathbf{p}}_1)Y_{\ell m}^*(\hat{\mathbf{p}}_2)
\end{align}
But we can decompose $\gamma(\mathbf{p}_1,\mathbf{p}_2)B(\mathbf{p}_1,\mathbf{p}_2)$ into Legendre polynomials, in particular the angle between $\mathbf{p}_1$ and $\mathbf{p}_2$
\begin{align}
\gamma(\mathbf{p}_1,\mathbf{p}_2)B(\mathbf{p}_1,\mathbf{p}_2) &=\sum_{\ell'}B_{\ell'}(p_1,p_2)P_{\ell'}(\hat{\mathbf{p}}_1\cdot \hat{\mathbf{p}}_2)\\
& = \sum_{\ell'}B_{\ell'}(p_1,p_2)\frac{4\pi}{2\ell'+1}\sum_{m'}Y_{\ell' m'}^*(\hat{\mathbf{p}}_1)Y_{\ell' m'}(\hat{\mathbf{p}}_2)
\end{align}
The integral becomes
\begin{align}
&(4\pi)^2\sum_{\substack{\ell,m\\ \ell',m'}}\frac{(-1)^{\ell}}{2\ell'+1} \int {\textnormal d}^3 p_1 {\textnormal d}^3 p_2  
B_{\ell'}(p_1,p_2)
M(p_1) M(p_2) j_{\ell}(p_1 r) j_{\ell}(p_2 r)\nonumber \\
& \qquad\qquad\qquad   Y_{\ell m}(\hat{\mathbf{p}}_1)Y_{\ell m}^*(\hat{\mathbf{p}}_2) Y_{\ell' m'}^*(\hat{\mathbf{p}}_1)Y_{\ell' m'}(\hat{\mathbf{p}}_2)
\end{align}
The integration over the angles  associated to $\mathbf{p}_1$ and $\mathbf{p}_2$ gives
\begin{align}
&(4\pi)^2\sum_{\ell,m}\frac{(-1)^{\ell}}{2\ell+1} \int {\textnormal d} p_1 {\textnormal d} p_2 p_1^2 p_2^2  
B_{\ell}(p_1,p_2)
M(p_1) M(p_2) j_{\ell}(p_1 r) j_{\ell}(p_2 r)\\
= & (4\pi)^2\sum_{\ell}(-1)^{\ell} \int {\textnormal d} p_1 {\textnormal d} p_2 p_1^2 p_2^2  
B_{\ell}(p_1,p_2)
M(p_1) M(p_2) j_{\ell}(p_1 r) j_{\ell}(p_2 r)
\end{align}
which gives
\begin{align}
& \int {\textnormal d}^3 p_1 {\textnormal d}^3 p_2 e^{i\mathbf{p}_{12}\cdot \mathbf{r}} \gamma(\mathbf{p}_1,\mathbf{p}_2) B(\mathbf{p}_1,\mathbf{p}_2) M(p_1) M(p_2) \nonumber\\
& =(4\pi)^2\sum_{\ell}{(-1)^{\ell}} \int {\textnormal d} p_1 {\textnormal d} p_2 p_1^2 p_2^2  
B_{\ell}(p_1,p_2)
M(p_1) M(p_2) j_{\ell}(p_1 r) j_{\ell}(p_2 r)
\end{align}
and therefore the integral (\ref{eq:Qeff}) becomes
\begin{align}
\frac{1}{4 \pi^4}\sum_{\ell}{(-1)^{\ell}} \int {\textnormal d} p_1 {\textnormal d} p_2 p_1^2 p_2^2  
B_{\ell}(p_1,p_2)
M(p_1) M(p_2) j_{\ell}(p_1 r) j_{\ell}(p_2 r)
\end{align}
with
\begin{align}
\gamma(\mathbf{p}_1,\mathbf{p}_2)B(\mathbf{p}_1,\mathbf{p}_2) =\sum_{\ell}B_{\ell}(p_1,p_2)P_{\ell}(\hat{\mathbf{p}}_1\cdot \hat{\mathbf{p}}_2)
\end{align}
and therefore
\begin{align}
B_\ell(p_1, p_2) = \frac{2\ell+1}{2} \int_{-1}^1 (1-x^2) B(p_1, p_2, p_3) P_\ell(x) {\textnormal d}x, \quad 
p_3 = \sqrt{p_1^2 + p_2^2 + 2p_1 p_2  x},
\end{align}

\section{Standard Perturbation Theory at 1-Loop}
\label{SPT1loop}
From \cite{Takushima:2013foa,Takushima:2015iha}, we obtain the following expressions at third order
\begin{align}
\delta_1(t, \mathbf{p}) & =D_{+}(t) \delta_{\mathrm{L}}(\mathbf{p}) \\
\Phi_1(t, \mathbf{p}) & =-\frac{a^2 H^2}{p^2} D_{+}(t) \kappa_{\Phi}(t) \delta_{\mathrm{L}}(\mathbf{p})\\
\delta_2(t, \mathbf{p}) &=D_{+}^2(t)\left(\mathcal{W}_\alpha(\mathbf{p})-\frac{2}{7} \lambda_1(t) \mathcal{W}_\gamma(\mathbf{p})\right) \\
\Phi_2(t, \mathbf{p}) &=-\frac{a^2 H^2}{p^2} D_{+}^2(t)\left(\kappa_{\Phi}(t) \mathcal{W}_\alpha(\mathbf{p})+\lambda_{\Phi}(t) \mathcal{W}_\gamma(\mathbf{p})\right),\\
\delta_3(t, \mathbf{p}) &=D_{+}^3(t)\left(\mathcal{W}_{\alpha \alpha}(\mathbf{p})-\frac{2}{7} \lambda_1(t) \mathcal{W}_{\alpha \gamma R}(\mathbf{p})-\frac{2}{7} \lambda_1(t) \mathcal{W}_{\alpha \gamma L}(\mathbf{p})-\frac{2}{21} \lambda_2(t) \mathcal{W}_{\gamma \gamma}(\mathbf{p})\right)\label{eq:AppB1}\\
\Phi_3(t, \mathbf{p}) &=-\frac{a^2 H^2}{p^2}\left(\kappa_{\Phi}(t) \delta_3(t, \mathbf{p})+D_{+}^3(t)\left(\sigma_{\Phi}(t) \mathcal{W}_{\gamma \alpha}(\mathbf{p})+\mu_{\Phi}(t) \mathcal{W}_{\gamma \gamma}(\mathbf{p})\right)\right),
\label{eq:AppB2}
\end{align}
Note that there is a coefficient in \( \delta_3 \) which we have omitted, as it does not contribute to the 1-loop correlation functions. The time-dependent coefficients are given by
\begin{align}
\kappa_\Phi & = \frac{3}{2}\Omega_m \mu F \\
\lambda_\Phi &=-\frac{3}{7}\lambda_1\Omega_m \mu F-\frac{3}{2}\Omega_m g (\mu F-1) \\
\label{sigmaphi}
\sigma_\Phi & = -3 g \Omega_m (\mu F-1) \\
\label{muphi}
\mu_\Phi &= 3 g \Bigl(g+\frac{2}{7}\lambda_1\Bigr)\Omega_m (\mu F-1)
\end{align}
with
\begin{align}
& \mathcal{W}_\alpha(\mathbf{p})=\frac{1}{(2 \pi)^3} \int d \mathbf{k}_1 d \mathbf{k}_2 \delta^{(3)}\left(\mathbf{k}_1+\mathbf{k}_2-\mathbf{p}\right) \alpha^{(s)}\left(\mathbf{k}_1, \mathbf{k}_2\right) \delta_{\mathrm{L}}\left(\mathbf{k}_1\right) \delta_{\mathrm{L}}\left(\mathbf{k}_2\right) \\
& \mathcal{W}_\gamma(\mathbf{p})=\frac{1}{(2 \pi)^3} \int d \mathbf{k}_1 d \mathbf{k}_2 \delta^{(3)}\left(\mathbf{k}_1+\mathbf{k}_2-\mathbf{p}\right) \gamma\left(\mathbf{k}_1, \mathbf{k}_2\right) \delta_{\mathrm{L}}\left(\mathbf{k}_1\right) \delta_{\mathrm{L}}\left(\mathbf{k}_2\right)\\
& \mathcal{W}_{\alpha \alpha}(\mathbf{p})=\frac{1}{(2 \pi)^6} \int d \mathbf{k}_1 d \mathbf{k}_2 d \mathbf{k}_3 \delta^{(3)}\left(\mathbf{k}_1+\mathbf{k}_2+\mathbf{k}_3-\mathbf{p}\right) \alpha \alpha\left(\mathbf{k}_1, \mathbf{k}_2, \mathbf{k}_3\right) \delta_L\left(\mathbf{k}_1\right) \delta_L\left(\mathbf{k}_2\right) \delta_L\left(\mathbf{k}_3\right),\\
& \mathcal{W}_{\alpha \gamma R}(\mathbf{p})=\frac{1}{(2 \pi)^6} \int d \mathbf{k}_1 d \mathbf{k}_2 d \mathbf{k}_3 \delta^{(3)}\left(\mathbf{k}_1+\mathbf{k}_2+\mathbf{k}_3-\mathbf{p}\right) \alpha \gamma_R\left(\mathbf{k}_1, \mathbf{k}_2, \mathbf{k}_3\right) \delta_L\left(\mathbf{k}_1\right) \delta_L\left(\mathbf{k}_2\right) \delta_L\left(\mathbf{k}_3\right)\\
& \mathcal{W}_{\alpha \gamma L}(\mathbf{p})=\frac{1}{(2 \pi)^6} \int d \mathbf{k}_1 d \mathbf{k}_2 d \mathbf{k}_3 \delta^{(3)}\left(\mathbf{k}_1+\mathbf{k}_2+\mathbf{k}_3-\mathbf{p}\right) \alpha \gamma_L\left(\mathbf{k}_1, \mathbf{k}_2, \mathbf{k}_3\right) \delta_L\left(\mathbf{k}_1\right) \delta_L\left(\mathbf{k}_2\right) \delta_L\left(\mathbf{k}_3\right)\\
& \mathcal{W}_{\gamma \gamma}(\mathbf{p}) \equiv \frac{1}{(2 \pi)^6} \int d \mathbf{k}_1 d \mathbf{k}_2 d \mathbf{k}_3 \delta^{(3)}\left(\mathbf{k}_1+\mathbf{k}_2+\mathbf{k}_3-\mathbf{p}\right) \gamma \gamma\left(\mathbf{k}_1, \mathbf{k}_2, \mathbf{k}_3\right) \delta_L\left(\mathbf{k}_1\right) \delta_L\left(\mathbf{k}_2\right) \delta_L\left(\mathbf{k}_3\right) \\
&\mathcal{W}_{\gamma \alpha}(\mathbf{p}) \equiv \frac{1}{(2 \pi)^6} \int d \mathbf{k}_1 d \mathbf{k}_2 d \mathbf{k}_3 \delta^{(3)}\left(\mathbf{k}_1+\mathbf{k}_2+\mathbf{k}_3-\mathbf{p}\right) \gamma \alpha\left(\mathbf{k}_1, \mathbf{k}_2, \mathbf{k}_3\right) \delta_L\left(\mathbf{k}_1\right) \delta_L\left(\mathbf{k}_2\right) \delta_L\left(\mathbf{k}_3\right)
\end{align}
and the kernels are given by
\cite{Takushima:2015iha}
\begin{align}
&\alpha \alpha\left(\mathbf{k}_1, \mathbf{k}_2, \mathbf{k}_3\right)  =\frac{1}{3}\left(\alpha^{(s)}\left(\mathbf{k}_1, \mathbf{k}_2+\mathbf{k}_3\right) \alpha^{(s)}\left(\mathbf{k}_2, \mathbf{k}_3\right)+2 \text { cyclic terms }\right)\\
&\alpha \gamma_R\left(\mathbf{k}_1, \mathbf{k}_2, \mathbf{k}_3\right) =\frac{1}{3}\left(\alpha\left(\mathbf{k}_1, \mathbf{k}_2+\mathbf{k}_3\right) \gamma\left(\mathbf{k}_2, \mathbf{k}_3\right)+2 \text { cyclic terms }\right),\\
&\alpha \gamma_L\left(\mathbf{k}_1, \mathbf{k}_2, \mathbf{k}_3\right)  =\frac{1}{3}\left(\alpha\left(\mathbf{k}_1+\mathbf{k}_2, \mathbf{k}_3\right) \gamma\left(\mathbf{k}_2, \mathbf{k}_3\right)+2 \text { cyclic terms }\right),\\
&\gamma \gamma\left(\mathbf{k}_1, \mathbf{k}_2, \mathbf{k}_3\right)  =\frac{1}{3}\left(\gamma\left(\mathbf{k}_1, \mathbf{k}_2+\mathbf{k}_3\right) \gamma\left(\mathbf{k}_2, \mathbf{k}_3\right)+2 \text { cyclic terms }\right)\\
&\gamma \alpha\left(\mathbf{k}_1, \mathbf{k}_2, \mathbf{k}_3\right) =\frac{1}{3}\left(\gamma\left(\mathbf{k}_1, \mathbf{k}_2+\mathbf{k}_3\right) \alpha^{(s)}\left(\mathbf{k}_2, \mathbf{k}_3\right)+2 \text { cyclic terms }\right)\\
&\alpha\left(\mathbf{k}_1, \mathbf{k}_2\right) =1+\frac{\mathbf{k}_1 \cdot \mathbf{k}_2}{k_1^2}\,,\quad
\alpha^{(s)}\left(\mathbf{k}_1, \mathbf{k}_2\right)  =1+\frac{\mathbf{k}_1 \cdot \mathbf{k}_2\left(k_1^2+k_2^2\right)}{2 k_1^2 k_2^2}\,,\quad 
\gamma\left(\mathbf{k}_1, \mathbf{k}_2\right)  =1-\frac{\left(\mathbf{k}_1 \cdot \mathbf{k}_2\right)^2}{k_1^2 k_2^2}
\end{align}
The time dependent functions $\lambda_1$ and $\lambda_2$ are given by
\begin{align}
\label{eq:lambda1}
\lambda_1(t) &=\frac{7}{2 D_{+}^2(t)} \int_0^t \frac{D_{-}(t) D_{+}\left(t^{\prime}\right)-D_{+}(t) D_{-}\left(t^{\prime}\right)}{W\left(t^{\prime}\right)} D_{+}^2\left(t^{\prime}\right)\left(f^2 H^2-N_\gamma\left(t^{\prime}\right)\right) d t^{\prime}\\
\label{eq:lambda2}
\lambda_2(t) &=-\frac{21}{2 D_{+}^3(t)} \int_0^t \frac{D_{-}(t) D_{+}\left(t^{\prime}\right)-D_{+}(t) D_{-}\left(t^{\prime}\right)}{W\left(t^{\prime}\right)} D_{+}^3\left(t^{\prime}\right) N_{\gamma \gamma}\left(t^{\prime}\right) d t^{\prime},
\end{align}
where $f$ is the linear growth rate $f(t) = d \ln D_+(t)/ d \ln a$, $W$ the Wronskian of the 2 independent solutions
\begin{align}
W(t) = D_+(t) \dot D_-(t) - \dot D_+(t)D_-(t) 
\end{align}
and
\begin{align}
    N_\gamma(t) &= -\frac{3}{2}g \Omega_m H^2 (\mu F-1)\\
    N_{\gamma \gamma}(t) & = 3g\Omega_m H^2 (\mu F-1)\Bigl(g-1+\frac{2}{7}\lambda_1\Bigr) -2 f^2 H^2+\frac{8}{7} f^2 H^2 \lambda_1+\frac{4}{7} f H \dot{\lambda}_1
\end{align}
The power spectrum $P_{\delta\delta}^{\text{1-loop}}$ has been computed in \cite{Takushima:2015iha}, where it was found to be given by $P_{\delta\delta}^{\text{1-loop}} = P_{\delta \delta}^{(22)} + 2 P_{\delta \delta}^{(13)}$
with
\begin{align}
&P_{\delta \delta}^{(22)}(t, p) =\frac{p^3}{98(2 \pi)^2} \int d r P_{\mathrm{L}}(r p) \int_{-1}^1 d x P_{\mathrm{L}}\left(p\left(1+r^2-2 r x\right)^{1 / 2}\right)  \frac{\left((7-4 \lambda_1) r+7 x+2(2 \lambda_1-7) r x^2\right)^2}{\left(1+r^2-2 r x\right)^2}\\
&2 P_{\delta \delta}^{(13)}(t, p) =  \frac{p^3}{252(2 \pi)^2} P_{\mathrm{L}}(p) \int d r P_{\mathrm{L}}(r p)\left[12 \lambda_2 \frac{1}{r^2}-2(21+36 \lambda_1+22 \lambda_2)+4(84-48 \lambda_1-11 \lambda_2) r^2\right. \\
& \qquad \left.-6(21-12 \lambda_1-2 \lambda_2) r^4+\frac{3}{r^3}\left(r^2-1\right)^3\left((21-12 \lambda_1-2 \lambda_2) r^2+2 \lambda_2\right) \ln \left(\frac{r+1}{|r-1|}\right)\right]
\end{align}
while we can obtain from Eqs. (\ref{eq:AppB1},\ref{eq:AppB2})  $P_{\Phi\delta}^{\text{1-loop}}=P_{\Phi\delta}^{22}+P_{\Phi\delta}^{13}+P_{\Phi\delta}^{31}$
with
\begin{align}
P_{\Phi\delta}^{22}(t,p) &= -\frac{a^2H^2}{p^2}\frac{2}{(2\pi)^3}\int d^3 k P_L(k)P_L(|\mathbf{p}-\mathbf{k}|)\Bigl[\kappa_\Phi \alpha^{(s)}(\mathbf{k},\mathbf{p}-\mathbf{k}) \alpha^{(s)}(\mathbf{k},\mathbf{p}-\mathbf{k})\nonumber\\
&+(\lambda_\Phi-\frac{2}{7}\lambda_1\kappa_\Phi) \alpha^{(s)}(\mathbf{k},\mathbf{p}-\mathbf{k}) \gamma(\mathbf{k},\mathbf{p}-\mathbf{k})-\frac{2}{7}\lambda_1\lambda_\Phi \gamma(\mathbf{k},\mathbf{p}-\mathbf{k}) \gamma(\mathbf{k},\mathbf{p}-\mathbf{k})\Bigr]\\
P_{\Phi\delta}^{13}(t,p) &= -\frac{a^2H^2}{p^2} \kappa_\Phi P_L(p)\frac{3}{(2\pi)^3}\int d^3 k P_L(k)\Bigl[\alpha\alpha(-\mathbf{k},\mathbf{k},\mathbf{p})
-\frac{2}{7} \lambda_1 \alpha\gamma_R(-\mathbf{k},\mathbf{k},\mathbf{p})
-\frac{2}{7} \lambda_1 \alpha \gamma_L(-\mathbf{k},\mathbf{k},\mathbf{p})\nonumber\\
&-\frac{2}{21}\lambda_2 \gamma \gamma(-\mathbf{k},\mathbf{k},\mathbf{p})\Bigr]\\
P_{\Phi\delta}^{31}(t,p) &= -\frac{a^2H^2}{p^2} P_L(p)\frac{3}{(2\pi)^3}\int d^3 k P_L(k)\Bigl[\kappa_{\Phi} \Bigl(\alpha\alpha(-\mathbf{k},\mathbf{k},\mathbf{p})
-\frac{2}{7} \lambda_1 \alpha\gamma_R(-\mathbf{k},\mathbf{k},\mathbf{p})
-\frac{2}{7} \lambda_1 \alpha \gamma_L(-\mathbf{k},\mathbf{k},\mathbf{p})\nonumber\\
&-\frac{2}{21} \lambda_2\gamma \gamma(-\mathbf{k},\mathbf{k},\mathbf{p})\Bigr)
+\sigma_{\Phi} \gamma\alpha(-\mathbf{k},\mathbf{k},\mathbf{p})
+\mu_{\Phi} \gamma\gamma(-\mathbf{k},\mathbf{k},\mathbf{p})\Bigr]
\end{align}

\end{document}